\documentclass[10pt,a4paper,english,superscriptaddress,floatfix,footnotesinbib]{revtex4-1}
\usepackage[T1]{fontenc}
\usepackage[latin9]{inputenc}
\usepackage{amsmath}
\usepackage{graphicx}
\usepackage{epstopdf}
\usepackage{etoolbox}

\makeatletter

\pdfpageheight\paperheight
\pdfpagewidth\paperwidth

 \@ifundefined{textcolor}{}
 {%
   \definecolor{BLACK}{gray}{0}
   \definecolor{WHITE}{gray}{1}
   \definecolor{RED}{rgb}{1,0,0}
   \definecolor{GREEN}{rgb}{0,1,0}
   \definecolor{BLUE}{rgb}{0,0,1}
   \definecolor{CYAN}{cmyk}{1,0,0,0}
   \definecolor{MAGENTA}{cmyk}{0,1,0,0}
   \definecolor{YELLOW}{cmyk}{0,0,1,0}
 }

\usepackage{bbm}
\usepackage{mathrsfs}
\usepackage{float}
\usepackage{mathtools}
\usepackage{cases}

\newcommand{\1}{\mathbbm 1}

\makeatother

\usepackage{babel}
\begin{document}

\title{Bayesian inference of epidemics on networks via Belief Propagation}

\author{Fabrizio Altarelli}
\affiliation{DISAT and Center for Computational Sciences, Politecnico di Torino,
Corso Duca degli Abruzzi 24, 10129 Torino, Italy}
\affiliation{Collegio Carlo Alberto, Via Real Collegio 30, 10024 Moncalieri, Italy}

\author{Alfredo Braunstein}
\affiliation{DISAT and Center for Computational Sciences, Politecnico di Torino,
Corso Duca degli Abruzzi 24, 10129 Torino, Italy}
\affiliation{Human Genetics Foundation, Via Nizza 52, 10126 Torino, Italy }
\affiliation{Collegio Carlo Alberto, Via Real Collegio 30, 10024 Moncalieri, Italy}

\author{Luca Dall'Asta}
\affiliation{DISAT and Center for Computational Sciences, Politecnico di Torino,
Corso Duca degli Abruzzi 24, 10129 Torino, Italy}
\affiliation{Collegio Carlo Alberto, Via Real Collegio 30, 10024 Moncalieri, Italy}

\author{Alejandro Lage-Castellanos}
\affiliation{DISAT and Center for Computational Sciences, Politecnico di Torino,
Corso Duca degli Abruzzi 24, 10129 Torino, Italy}
\affiliation{Physics Faculty, Havana University, San Lazaro y L, 10400 Habana, Cuba}

\author{Riccardo Zecchina}
\affiliation{DISAT and Center for Computational Sciences, Politecnico di Torino,
Corso Duca degli Abruzzi 24, 10129 Torino, Italy}
\affiliation{Human Genetics Foundation, Via Nizza 52, 10126 Torino, Italy }
\affiliation{Collegio Carlo Alberto, Via Real Collegio 30, 10024 Moncalieri, Italy}

\begin{abstract}
We study several bayesian inference problems for irreversible stochastic epidemic models on networks from a statistical physics viewpoint. We derive equations which allow to accurately compute the posterior distribution of the time evolution of the state of each node given some observations. At difference with most existing methods, we allow very general observation models, including unobserved nodes, state observations made at different or unknown times, and observations of infection times, possibly mixed together. Our method, which is based on the Belief Propagation algorithm, is efficient, naturally distributed, and exact on trees. As a particular case, we consider the problem of finding the ``zero patient'' of a SIR or SI epidemic given a snapshot of the state of the network at a later unknown time. Numerical simulations show that our method outperforms previous ones on both synthetic and real networks, often by a very large margin.
\end{abstract}

\maketitle

Tracing epidemic outbreaks in order to pin down their origin is a paramount problem in epidemiology. Compared to the pioneering work of John Snow on $1854$ London's cholera hit \cite{snow_mode_1855}, modern computational epidemiology can rely on accurate clinical data and on powerful computers to run large-scale simulations of stochastic compartment models. However, like most \emph{inverse} epidemic problems, identifying the origin (or {\em seed}) of an epidemic outbreak remains a challenging problem even for simple stochastic epidemic models, such as the susceptible-infected (SI) model and the susceptible-infected-recovered (SIR) model.

Several studies have recently proposed maximum likelihood estimators based on various kinds of information:  topological centrality \citep{shah_detecting_2010, shah_rumors_2011, comin_identifying_2011}, measures of the distance between observed data and the typical outcome of propagations from given initial conditions \citep{antulov-fantulin_statistical_2013}, or the estimation of the single most probable path \citep{zhu_information_2013}. Other estimators are derived under strong simplifying assumptions on the graph structure or on the spreading process \citep{luo_identifying_2013, pinto_locating_2012}. Notably, for a continuous time diffusion model with gaussian transmission delays, the estimator in \citep{pinto_locating_2012} is optimal for trees.
An interesting  systematic approach is Dynamic Message Passing (DMP) \citep{lokhov_inferring_2013}, which consists in a direct approximation and maximization of the likelihood function. DMP makes use of an approximate description of the stochastic process, inspired by statistical physics and relying on some decorrelation assumption, which is very accurate in providing local probability marginals \citep{karrer_message_2010}. However, as noted by the authors, it has two drawbacks. First, the space of initial conditions considered must be explored exhaustively. Second, DMP relies on a further assumption of single-site factorization of the likelihood function which is not necessarily consistent with the more accurate underlying approximation in \citep{karrer_message_2010}.

In this Letter we derive the Belief Propagation (BP) equations for the probability distribution of the time evolution of the state of the system conditioned on some observations. BP only relies on a decorrelation assumption similar to the one of \citep{karrer_message_2010}, and is therefore exact on trees. Extensive numerical simulations show that it is typically a very good approximation on general graphs. BP can be used to identify the origin of an epidemic outbreak in the SIR, SI, and similar models, even with multiple infection seeds and incomplete or heterogeneous information.

\paragraph*{The SIR model on graphs\label{sec:The-SIR-model-on-graphs}.}

We consider the susceptible-infected-recovered (SIR) model of spreading, a stochastic dynamical model in discrete time defined over a graph $G=(V,E)$. At time $t$ a node $i$ can be in one of three states 
represented by a variable $x_{i}^{t}\in\left\{ S,I,R\right\} $.
At each time step $t$, each infected node $i$ infects each one of its susceptible neighbors $\left\{ j\in\partial i:x_{j}^{t}=S\right\}$ with independent probabilities $\lambda_{ij}\in[0,1]$; then, node $i$ recovers with probability $\mu_{i}\in\left[0,1\right]$.
The dynamics is irreversible, as a given node can only undergo the transitions $S\to I\to R$.
Two important special cases of SIR are the independent cascades model (obtained when $\mu_{i}\equiv1$) \citep{kempe_maximizing_2003} and the susceptible-infected model (obtained when $\mu_{i}\equiv0$).

\paragraph*{SIR dynamics as a graphical model and Bayesian inference.}

Assume that a certain set of nodes initiates the infection at time $t=0$,
i.e. with $x_{i}^{0}=I$. A realization of the SIR process can be univocally expressed in terms of a set of independent {\em recovery times} $g_{i}$, distributed as $\mathcal{G}_{i}\left(g_{i}\right)=\mu_{i}\left(1-\mu_{i}\right)^{g_{i}}$,
and conditionally independent {\em transmission delays} $s_{ij}$ following a geometric distribution  $\omega_{ij}\left(s_{ij}|g_{i}\right)=\lambda_{ij}\left(1-\lambda_{ij}\right)^{s_{ij}}$ for $s_{ij}\leq g_i$ and $\omega_{ij}\left(\infty|g_i\right)=\sum_{s>g_i}\lambda_{ij}\left(1-\lambda_{ij}\right)^{s}$.
The infection times $t_{i}$ can be deterministically computed by imposing the condition $1=\phi_{i}=\delta\left(t_{i},\1\left[x_{i}^{0}= S\right]\left(\min_{j\in\partial i}\left\{ t_{j}+s_{ji}\right\} +1\right)\right)$ for every $i$.

The distribution of infection and recovery times $\mathbf{t},\mathbf{g}$
given the initial state $\mathbf{x}^{0}$ can thus be written as
\begin{eqnarray}
\nonumber \mathcal{P}\left(\mathbf{t},\mathbf{g}|\mathbf{x}^{0}\right)  & = & \sum_{\mathbf{s}}\mathcal{P}\left(\mathbf{t}|\mathbf{x}^{0},\mathbf{g},\mathbf{s}\right)\mathcal{P}\left(\mathbf{s}|\mathbf{g}\right)\mathcal{P\left(\mathbf{g}\right)}\\
& = &  \sum_{\mathbf{s}}\prod_{i,j}\omega_{ij}\prod_{i}\phi_{i}\mathcal{G}_{i}.
	\label{eq:direct}
\end{eqnarray}
In the inference problem we initially assume that (i) at time $t=T$ the state of every node $x_{i}^{T}\in\left\{ S,I,R\right\}$ is known, and (ii) at $t=0$ the state of every node was extracted independently from the prior distribution  $\gamma_i(x_i^0) = \gamma \1[x^0_i=I] + (1-\gamma) \1[x^0_i=S]$.
Using Bayes' theorem and \eqref{eq:direct}, the posterior can be expressed as
\begin{eqnarray}
\nonumber \mathcal{P}\left(\mathbf{x}^{0}|\mathbf{x}^{T}\right) & \propto & \sum_{\mathbf{t,g}}\mathcal{P}\left(\mathbf{x}^{T}|\mathbf{t},\mathbf{g}\right)\mathcal{P}\left(\mathbf{t},\mathbf{g}|\mathbf{x}^{0}\right)\mathcal{P}\left(\mathbf{x}^{0}\right) \\
& = & \sum_{\mathbf{t},\mathbf{g},\mathbf{s}}\prod_{i,j}\omega_{ij}\prod_{i}\phi_{i}\mathcal{G}_{i}\gamma_{i}\zeta_i \label{eq:Bayes}
\end{eqnarray}
where we exploited the fact that the state $\mathbf{x}^T$ at time $T$ given a set $(\mathbf{t},\mathbf{g})$ of infection and recovery times follows a deterministic law $\mathcal{P}\left(\mathbf{x}^{T}|\mathbf{t},\mathbf{g}\right)=\prod_{i}\zeta_i\left(t_{i},g_{i},x_{i}^{T}\right)$
where $\zeta_i=\1\left[x_{i}^{T}=S,T < t_i\right]+\1\left[x_{i}^{T}=I,t_{i}\leq T<t_{i}+g_{i}\right]+\1\left[x_{i}^{T}=R,t_{i}+g_{i}\leq T\right]$.

\paragraph*{Belief Propagation equations for the posterior.}

Finding the marginals of \eqref{eq:Bayes} is computationally hard, and we propose to approximate them using BP. We will borrow from graphical models the {\em factor graph} representation of the dependence of the factors on their variables in a generalized Boltzmann distribution. It is convenient to introduce the new variables $t_{j}'=t_{j}+s_{ji}$ and eliminate the $s_{ij}$ and $s_{ji}$ parameters from the graphical model. By
defining the factors $\phi_{ij}=\omega_{ij}\left(t'_{i}-t_{i}|g_{i}\right)\omega_{ji}\left(t'_{j}-t_{j}|g_{j}\right)$ and $\phi_{i}=\delta\left(t_{i}, \1\left[x_{i}^{0}= S\right]\left(\min_{j\in\partial i}\left\{ t'_{j}\right\} +1\right)\right)$, the posterior becomes $\mathcal{P}\left(\mathbf{x}^{0}|\mathbf{x}^{T}\right)\propto\sum_{\mathbf{t},\mathbf{t}',\mathbf{g}}\prod_{i<j}\phi_{ij}\prod_{i}\phi_{i}\mathcal{G}_{i}\gamma_{i}\zeta_i$.

\begin{figure}
\includegraphics[width=0.5\columnwidth]{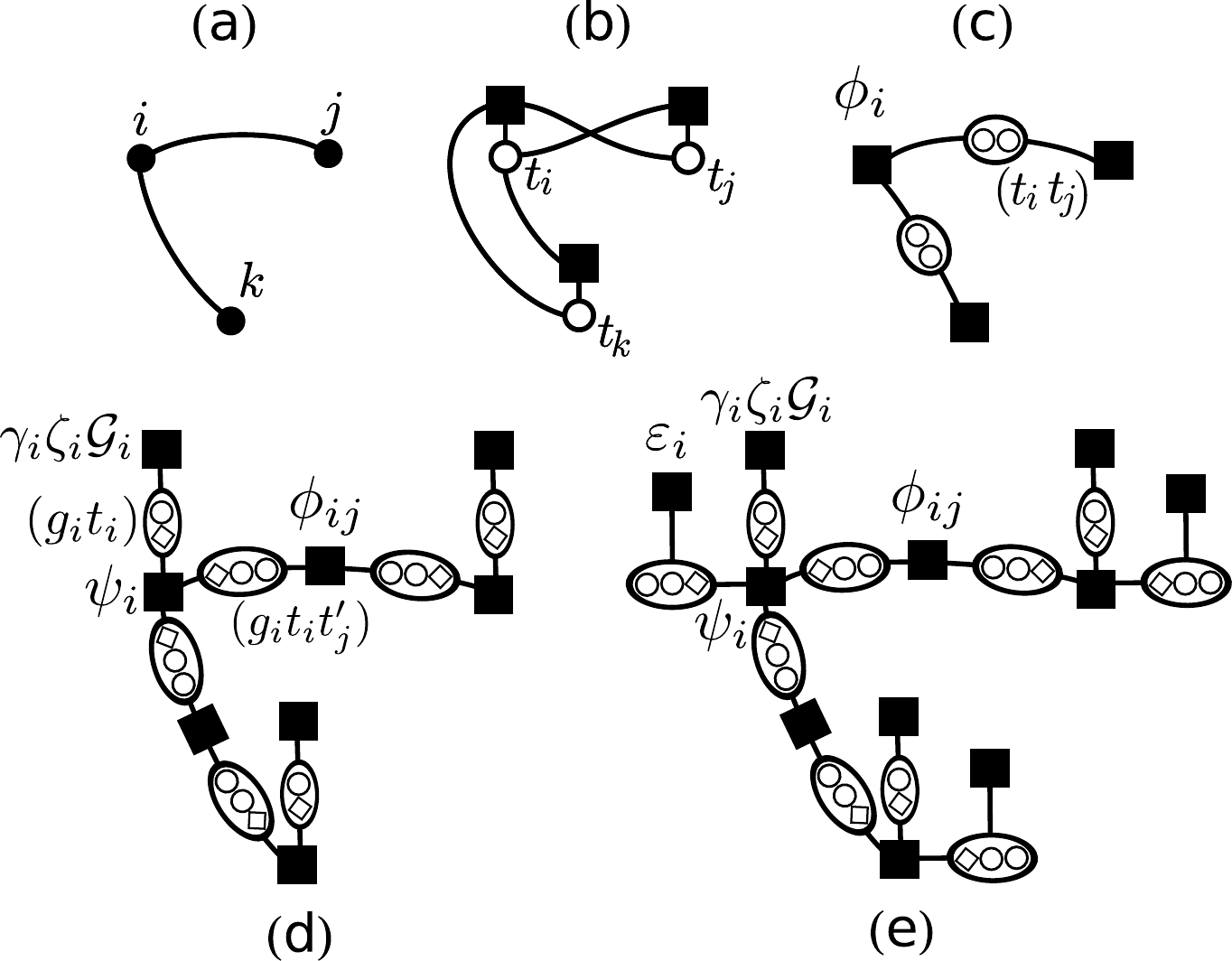}
\caption{Factor graph representations for irreversible dynamics\label{fig:fg}: full squares represent the factors of a generalized Boltzman distribution and ellipses the variables on which they depend. (a) Original graph (b) Loopy, naive factor graph for a deterministic dynamics (c) Disentangled dual tree factor graph (d) Factor graph for the SIR model given in \eqref{eq:effective} with known epidemic age (e) Factor graph representation with unknown epidemic age.}
\end{figure}

Note that even in the simple deterministic case $\mu_{i}\equiv\lambda_{ij}\equiv1$ where $t'_i = t_i$, the graphical model corresponding to the factors $\phi_i$ has the loopy representation displayed in Fig.\ref{fig:fg}b. The representation can be disentangled (see also \citep{altarelli_optimizing_2013,*altarelli_large_2013}) by grouping pairs of activation times $(t_i,t_j)$ in the same variable node (see Fig.\ref{fig:fg}c), which is crucial to make the BP approximation more accurate, and exact
on trees.
Similarly for the general case \eqref{eq:Bayes}, we introduce the triplets $(g_{i}^{(j)},t_{i}^{(j)},t'_{j})$
and group the constraints $\phi_i$ with compatibility checks into the factor node $\psi_{i}=\phi_{i}(t_{i},\mathbf{t}'_{\partial i})\prod_{j\in\partial i}\delta(t_{i}^{(j)},t_{i})\delta(g_{i}^{(j)},g_{i})$ (See Fig.~\ref{fig:fg}d) to obtain an {\em effective} model
\begin{equation}
\mathcal{Q}=\frac1Z \prod_{i<j}\phi_{ij}\prod_{i}\psi_{i}\mathcal{G}_{i}\gamma_{i}\zeta_i
\label{eq:effective}
\end{equation}
so that $\mathcal{P}\left(\mathbf{x}^{0}|\mathbf{x}^{T}\right)\propto\sum_{\mathbf{t},\mathbf{t}',\mathbf{g}}\mathcal{Q}\left(\mathbf{g},\mathbf{t},\mathbf{t}',\mathbf{x}^{0}\right)$.
As the topology of the factor graph now mirrors the one of the original network, this approach allows the exact computation of posterior marginals for the SIR model on acyclic graphs.

We derived the BP equations for \eqref{eq:effective} (See appendix). A single BP iteration can be computed in time $O\left(T\cdot G^{2}\cdot\left|E\right|\right)$, where $G$ is the maximum allowed recovery time, which can be assumed constant for a geometric distribution $\mathcal G$. Once the BP equations converge, estimates for the infection time $t_i$ of each node are obtained, and nodes can be ranked by the posterior probability of being a seed of the epidemics $\mathcal{P}(x_i^0=I|\mathbf{x}^T)$ in decreasing order.

\begin{figure}
\includegraphics[width=0.7\columnwidth, angle=0]{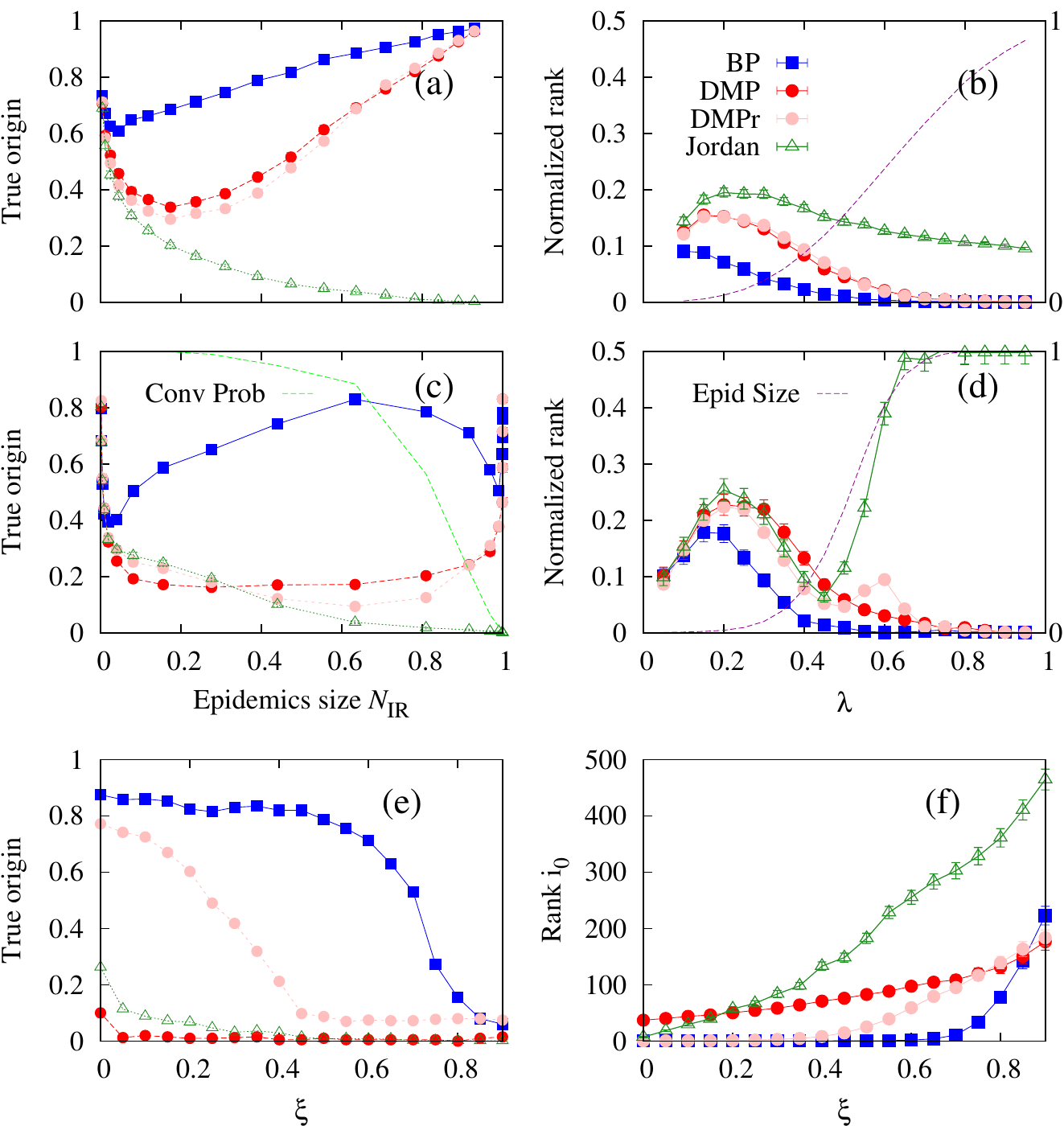}
\caption{\label{fig:rank_vs_lambda_T11_mu05}(Color online) Comparison between BP and DMP, DMPr and Jordan methods. Each point is an averge over $1000$ epidemics in random graphs with $N=1000$. Panels (a)--(b) are for SFG with average degree $\left< k \right> = 4$, recovery probability $\mu=0.5$ and observation time $T=5$. Panels (c)--(f) are for RRG with $k=4$, $\mu=0.5$ and $T=10$.
Panels (a) and (c): probability of finding the true origin $i_0$ of the epidemics as a function of the average epidemics size $N_{\mathrm{IR}}$. Panels (b) and (d): normalized rank of the true origin $(\mathrm{rank}\; i_0)/ N_{\mathrm{IR}}$ as a function of the transmission probability $\lambda$. The normalized epidemics size $N_{\mathrm{IR}}/N$ is also plotted (purple, right axis) versus $\lambda$. For very large epidemics BP may fail to converge in (a) and (b) within the specified number of iterations (the convergence probability is plotted in green in (a)), but relevant information is still present in the (unconverged) marginals. Panel (e): probability of finding the true origin of the epidemic as a function of the fraction of unobserved sites $\xi$ for transmission probability $\lambda=0.5$ and recovery probability $\mu=1$. Panel (f): Absolute rank given by each algorithm to the true origin as a function of $\xi$. \label{fig:rank_vs_obs}}
\end{figure}

\paragraph*{Identification of a single seed.}
We compared the inference performance of BP, of DMP, of a DMP variant we call DMPr (see appendix) and of the Jordan centrality method~\cite{shah_detecting_2010,shah_rumors_2011,comin_identifying_2011} on random graphs. We considered random regular graphs (RRG) with degree $k=4$ and preferential attachment scale-free graphs (SFG) with average degree $\left<k\right>=4$, both with $N=1000$ nodes and homogeneous propagation probability $\lambda_{ij}\equiv \lambda$ and recovery probability $\mu_i\equiv \mu$. Simulations summarized in Fig.~\ref{fig:rank_vs_lambda_T11_mu05}a--\ref{fig:rank_vs_lambda_T11_mu05}d show that BP generally outperforms the other methods by a large margin (see appendix).

\paragraph*{Incomplete information.}

In a more realistic setup, much of the information we assumed to know can be missing. First, a fraction $\xi$ of the nodes might be unobserved.
Figs.~\ref{fig:rank_vs_obs}e,\ref{fig:rank_vs_obs}f show the performance of the four methods in this case.
BP finds the true origin in more than $70\%$ of the instances with up to $\xi=60\%$, and it outperforms the other three methods for almost all $\xi$.

\begin{figure*}
\includegraphics[height=0.49\textwidth, angle=270]{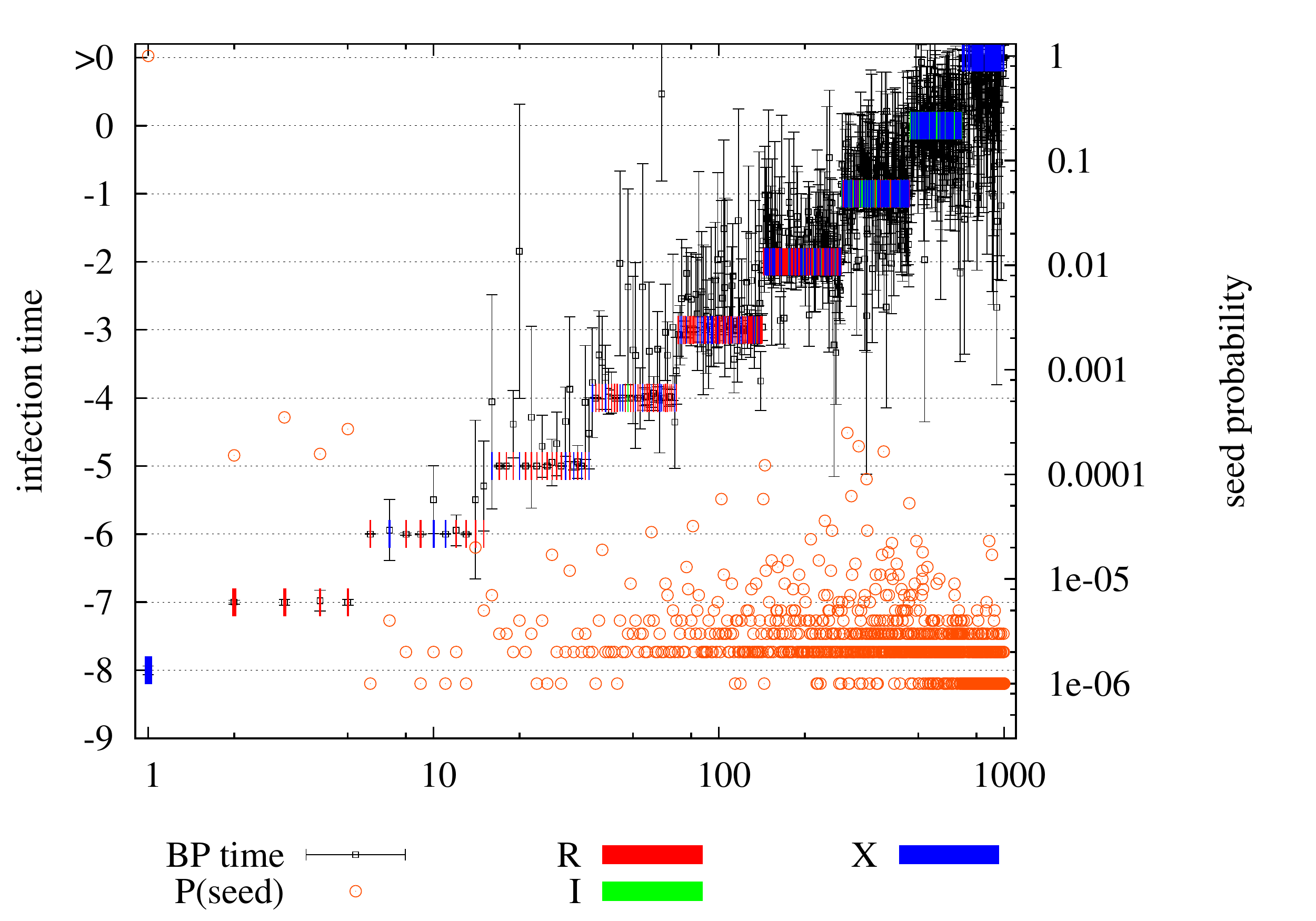}
\includegraphics[height=0.49\textwidth,angle=270]{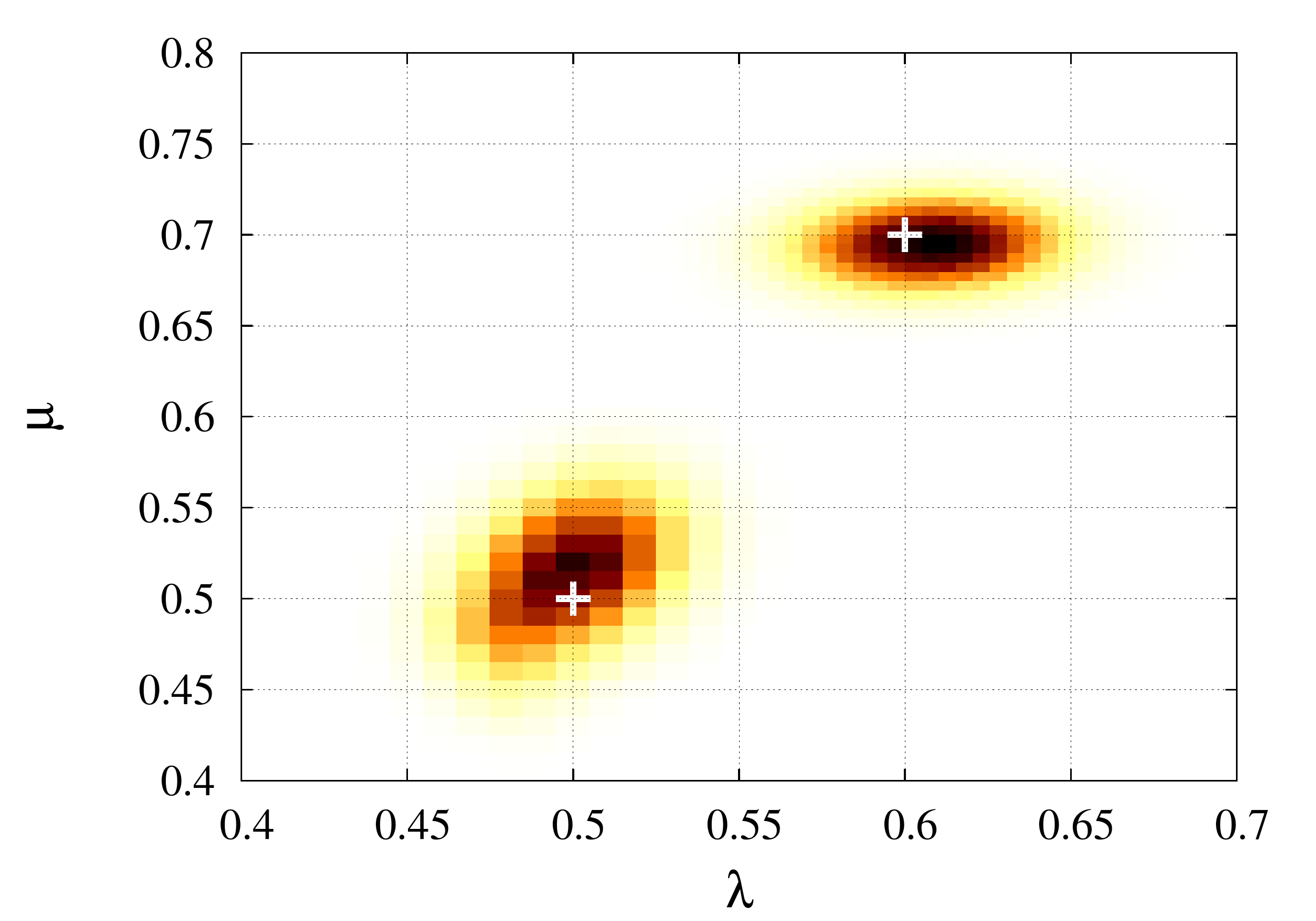}
\caption{(Color online) Examples of inference with incomplete information on a RRG with $N=1000$, $k=4$. Left: inference with $\lambda=0.7$, $\mu=0.6$, $\gamma=10^{-6}$, from observations fraction $1-\xi=0.6$ at time $T=0$ for an epidemics with $T_0 = -8$ (unknown to BP). Each colored block (R: recovered, I: infected and X: unknown) corresponds to a vertex, ordered in the horizontal axis by its real infection time given in the vertical axis. The mean and standard deviation of their BP posterior marginal distribution of infection time is plotted (black dots and error bars) along with the marginal posterior probability of spontaneous infection (orange, circles, right axis).\label{fig:infertimes} Right: Inference of epidemic parameters.
Heat-plot of the likelihood density of the parameters for two virtual epidemics. The first one with $\lambda=0.7$, $\mu=0.6$, $\Delta T=8$ (size $N_{\mathrm{IR}}=653$) shows a maximum of the estimated likelihood at $\hat \lambda=0.695$ and $\hat \mu=0.605$, and the second with $\lambda=0.5$, $\mu=0.5$, $\Delta T=10$ (size $N_{\mathrm{IR}}=462$) shows a maximum at $\hat \lambda=0.5$ and $\hat \mu=0.52$.
\label{fig:parameters}}
\end{figure*}

Second, the initial time $T_0$ and thus the age $\Delta T=T-T_0$ of the epidemics could be unknown. For a given upper bound on $\Delta T$, it suffices to consider the dynamical process to start from the all-susceptible state but to allow nodes to be spontaneously infected at an arbitrary time. This is equivalent to the addition of a fictitious neighbor to every node with no constraint $\psi_i$ in its activation time but with a prior probability $\varepsilon_i(g''_i,t''_i,t_i)=\delta(t''_i,\infty)(1-\gamma)+[1-\delta(t''_i,\infty)]\gamma$ of spontaneous infection (See Fig.~\ref{fig:fg}e). An example of inference for an epidemics with transmission and recovery probabilities $\lambda=0.7$ and $\mu=0.6$ is shown in Fig.~\ref{fig:infertimes}. The plot shows large correlation between true and inferred infection times, and also that the true origin (which was not observed) corresponds to the individual with largest inferred probability.

Finally, the proposed approach can be also used to estimate the epidemic parameters. Indeed, the partition function $Z$ in \eqref{eq:effective} is proportional to the likelihood of the unknown parameters. The log-likelihood $\log Z$ is well-approximated by the opposite of the Bethe free energy, which can be computed easily as a function of the BP messages at the fixed point. We show results for two different realizations of epidemics in Fig.~\ref{fig:parameters}. BP equations are iterated for equally spaced parameters  $\mu$ and $\lambda$ in $[0,1]$, and the Bethe free energy is computed after convergence. In both cases the epidemic age and the origin are correctly inferred and the parameters are recovered with good accuracy. In a real setting the search for the point of maximum likelihood can be performed with an expectation-maximization scheme rather than with an exhaustive search.

\begin{figure}
\includegraphics[width=0.7\columnwidth, angle=0]{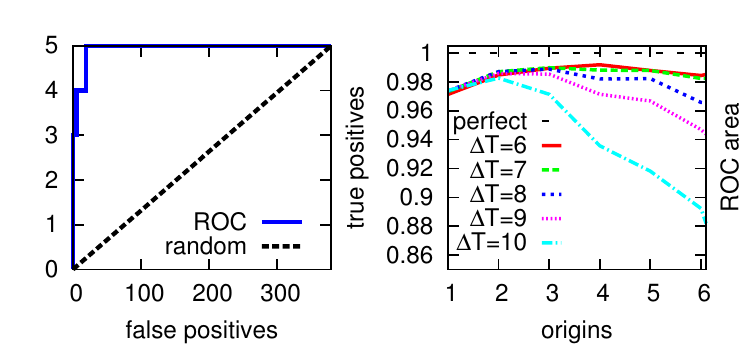}
\caption{(Color online) BP inference of multiple seeds for virtual epidemics on a RRG with $N=1000$, $k=4$ and $\mu=1$, $ 	\lambda=0.5$, $\gamma=0.002$ and different epidemic ages $\Delta T$ (unknown to BP). Left: an example {\em receiver operating characteristics} (ROC) curve for a sample with 5 origins and $\Delta T=8$, $N_{\mathrm{IR}}=379$. The first three ranked nodes are true seeds, and the average normalized area below the curve is 0.975. Right: the average normalized area below the ROC curve vs. the number of true origins on 1000 samples. Each ROC curve was computed restricted to the epidemic subgraph.\label{fig:multiseeds} }
\end{figure}

\paragraph*{Multiple seeds.}
If the epidemics initiates at multiple seeds, methods based on the exhaustive exploration of initial states like DMP suffer a combinatorial explosion. This problem does not affect BP, as the trace over initial conditions is performed directly within the framework. Fig.~\ref{fig:multiseeds} displays experiments with multiple seeds on RRG, showing that effective inference can also be achieved in this regime (see appendix).

\paragraph*{Evolving networks.}
We studied the case of time-dependent transmission probabilities $\lambda_{ij}$. This scenario can be analyzed by considering a distribution of transmission delays $\omega_{ij}(s_{ij}|g_i,t_i)$ depending explicitly on infection times $t_i$ (see appendix). We considered two real-world datasets of time-stamped contacts between pairs of individuals, which we aggregated into $\Delta T$ effective time steps.
(\emph{i}) A dataset describing $20$ seconds face-to-face contacts in an exhibition \cite{isella_whats_2011}. We employed the following parameters: probability of contagion in a 20 second interval $\lambda^{20s}_{ij}=0.2$, recovery probability $\mu^{20s}=0.0014$. (\emph{ii}) A dataset of sexual encounters self-reported on a website \cite{rocha_information_2010}. We set the probability of transmission in a single contact as $\lambda^{contact}_{ij}=0.2$ (within the range considered in \cite{rocha_information_2010}) and choose $\mu^{year}=0.5$. Results on the inference of simulated epidemics on both datasets are summarized in Table \ref{tab:real} showing a striking difference in favour of BP (see full results in the appendix). We tried to determine if the performance of the inference process was favored or hindered by temporal and spatial correlations present in the dataset, that are known to affect significantly the size of the outbreak \cite{rocha_simulated_2011} in some cases. However, after destroying the correlations in  one of such cases, we found that the performance of BP was essentially unchanged (see appendix).

\begin{table}
\setlength{\tabcolsep}{0.15cm}
\begin{tabular}{rcc|cccc}
	 & \multicolumn{2}{c}{Proximity} & \multicolumn{4}{c}{Sexual}\\
	$\Delta T$ & $20$ & $30$ & $5$ & $10$ & $15$ & $20$\\
	samples & 4\,082 & 1\,649 & 2\,636 & 2\,611 & 2\,592 & 2\,597\\
	\hline
	BP & 64/1.4 & 58/2.2 & 91/0.1 & 83/0.2 & 80/0.2 & 78/0.3\\
	DMPr & 60/1.6 & 54/2.2 & 79/0.2 & 61/0.6 & 58/0.9 & 55/1.2\\
	DMP & 56/1.8 & 53/2.4 & 79/0.2 & 59/0.9 & 55/1.6 & 53/2.4\\
\end{tabular}
\caption{Summary of results for virtual epidemics involving 10 or more individuals on real evolving networks. Each cell is in the format $p$/$r_{0}$, where $p$ is the probability of perfect inference (\%) and $r_0$ is the average ranking given to the real zero-patient.\label{tab:real}}
\end{table}

\paragraph*{Conclusions.}
We introduced a systematic, consistent and computationally efficient approach to the calculation of posterior distributions and likelihood of model parameters for a broad class of epidemic models. Besides providing the exact solution for acyclic graphs, we have shown the approach to be extremely effective also for synthetic and real networks with cycles, both in a static and a dynamic context. More general epidemic models such as the Reed-Frost model \citep{bailey_mathematical_1975} that include latency and incubation times, and other observation models \citep{zhu_information_2013, pinto_locating_2012} can be analyzed with a straightforward generalization by simply defining appropriate recovery $\mu_{i}$ and transmission probabilities $\lambda_{ij}$ that depend on the time after infection and by employing modified observation laws $\zeta_i$.

\paragraph*{Acknowledgments.}The authors acknowledge the European Grants FET Open No. 265496 and ERC No. 267915 and Italian FIRB Project No. RBFR10QUW4.
\AtEndEnvironment{thebibliography}{
}
\appendix

\section{Belief propagation equations}

We show the derivation of the Belief Propagation equations (also known as {\em replica-symmetric cavity equations} in statistical physics) for the posterior distribution of the SIR model

\begin{equation}
\mathcal{Q}=\frac1Z\prod_{i<j}\phi_{ij}\prod_{i}\psi_{i}\gamma_{i}\zeta_{i}\mathcal{G}_{i}\label{eq:effective}
\end{equation}

where

\begin{eqnarray}
\psi_{i}\left(t_{i},g_{i},\left\{ \left(t_{i}^{(j)},t_{j}^{'(j)},g_{i}^{(j)}\right)\right\} _{j\in\partial i}\right) & = & \left\{\delta\left(t_{i},0\right)+\delta\left[t_{i},\left(1+\min_{j\in\partial i}\left\{ t_{j}^{'\left(j\right)}\right\} \right)\right]\right\}\prod_{j\in\partial i}\delta\left(g_{i},g_{i}^{\left(j\right)}\right)\delta\left(t_{i},t_{i}^{\left(j\right)}\right)\\
\phi_{ij}\left(\left(t_{i},t_{j}^{'},g_{i}\right),\left(t_{j},t_{i}^{'},g_{j}\right)\right) & = & \omega_{ij}\left(t'_{i}-t_{i}|g_{i}\right)\omega_{ji}\left(t'_{j}-t_{j}|g_{j}\right)\\
\gamma_{i}\left(t_{i}\right) & = & \gamma\delta\left(t_{i},0\right)+\left(1-\gamma\right)\left(1-\delta\left(t_{i},0\right)\right)\\
\zeta_{i}\left(t_{i},g_{i},x_{i}^{T}\right) & = & \1\left[x_{i}^{T}=I,t_{i}\leq T<t_{i}+g_{i}\right]+\1\left[x_{i}^{T}=S,T<t_{i}\right]+\\
 &  & +\1\left[x_{i}^{T}=R,t_{i}+g_{i}\leq T\right]\\
\mathcal{G}_{i}\left(g_{i}\right) & = & r_{i}\left(1-r_{i}\right)^{g_{i}}
\end{eqnarray}

Belief propagation consists in a set of equations for
single-site probability distributions labeled by directed graph edges. These equations are solved by iteration, and on a fixed point give an approximation for single-site marginals and other quantities of interest like the partition function $Z$.

We recall the general form of the BP equations in the following. For a factorized probability measure on $\underline{z}=\{z_i\}$,
\begin{equation}
	M(\underline{z}) = \frac1Z \prod_a F_a(\underline{z}_a)
\end{equation}
where $\underline{z}_a$ is the subvector of variables that $F_a$ depends on, the general form of the equations is
\begin{eqnarray}
	p_{F_{a}\to i}\left(z_{i}\right) & = & \frac{1}{Z_{ai}} \sum_{\left\{ z_{j}:j\in \partial a\setminus i\right\} }F_{a}\left(\left\{ z_{i}\right\} _{i\in\partial a}\right)\prod_{j\in\partial a\setminus i}m_{j\to F_a}\left(z_{j}\right)\label{eq:factor-to-var}\\
	m_{i\to F_{a}}\left(z_{i}\right) & = & \frac{1}{Z_{ia}} \prod_{b\in \partial i\setminus a}p_{F_b\to i}\left(z_{i}\right)\label{eq:var-to-factor}\\
	m_{i}\left(z_{i}\right) & = & \frac{1}{Z_{i}} \prod_{b\in \partial i}p_{F_b\to i}\left(z_{i}\right)\label{eq:var-marginal}
\end{eqnarray}
where $F_{a}$ is a {\em factor} (i.e. $\psi_{i}$, $\phi_{ij}$, $\gamma_{i}$, $\zeta_{i}$
or $\mathcal{G}_{i}$ in our case), $z_{i}$ is a variable (i.e.
$(t_{i},g_{i})$,$(t_{i}^{\left(j\right)},t_{j}^{'\left(j\right)},g_{i}^{\left(j\right)})$
in our case), $\partial a$ is the subset of indices of variables in factor $F_a$ and $\partial i$ is the subset of factors that depend on $z_i$. Terms $Z_{ia},Z_{ai}$ and $Z_i$ are normalization factors that can be calculated once the rest of the right-hand side is computed. While equations \eqref{eq:var-to-factor}-\eqref{eq:var-marginal} can be always computed
efficiently in general, the computation of the trace in \eqref{eq:factor-to-var} may need a time which is exponential
in the number of participating variables. The update equations \eqref{eq:factor-to-var}
for factors $\phi_{ij}$, $\gamma_{i}$, $\zeta_{i}$ and  $\mathcal{G}_{i}$ can be computed in a straightforward way because they involve a very small (constant) number of variables each.
We show the derivation of an efficient version of equation \eqref{eq:factor-to-var} for factor $\psi_{i}$ that can be computed in linear time in the degree of vertex $i$:{\small
\begin{eqnarray}
p_{\psi_{i}\to j}\left(t_{i}^{\left(j\right)},t_{j}^{'\left(j\right)},g_{i}^{\left(j\right)}\right) & \propto & \sum_{g_{i},t_{i}}\sum_{\left\{ t_{i}^{\left(k\right)},t_{k}^{'\left(k\right)},g_{i}^{\left(k\right)}\right\} }m_{i\to\psi_{i}}\left(t_{i},g_{i}\right)\times\\
 &  & \times\prod_{k\in\partial i\setminus j}m_{k\to\psi_{i}}\left(t_{i}^{\left(k\right)},t_{k}^{'\left(k\right)},g_{i}^{\left(k\right)}\right)\psi_{i}\left(t_{i},g_{i},\left\{ \left(t_{i}^{(k)},t_{k}^{'(k)},g_{i}^{(k)}\right)\right\} _{k\in\partial i}\right)\nonumber \\
 & \propto & m_{i\to\psi_{i}}\left(t_{i}^{\left(j\right)},g_{i}^{\left(j\right)}\right)\sum_{\left\{ t_{k}^{'\left(k\right)}\right\} }\prod_{k\in\partial i\setminus j}m_{k\to\psi_{i}}\left(t_{i}^{\left(j\right)},t_{k}^{'\left(k\right)},g_{i}^{\left(j\right)}\right)\times\\
 &  & \times\left[\delta\left(t_{i}^{(j)},0\right)+\delta\left(t_{i}^{(j)},\left(1+\min_{k\in\partial i}\left\{ t_{k}^{'\left(k\right)}\right\} \right)\right)\right]\nonumber \\
 & \propto & \delta\left(t_{i}^{\left(j\right)},0\right)m_{i\to\psi_{i}}\left(0,g_{i}^{\left(j\right)}\right)\prod_{k\in\partial i\setminus j}\sum_{t_{k}^{'\left(k\right)}}m_{k\to\psi_{i}}\left(0,t_{k}^{'\left(k\right)},g_{i}^{\left(j\right)}\right)+\label{eq:pass}\\
 &  & + \; m_{i\to\psi_{i}}\left(t_{i}^{\left(j\right)},g_{i}^{\left(j\right)}\right)\1\left(t_{i}^{\left(j\right)}\leq t_{j}^{'\left(j\right)}+1\right)\prod_{k\in\partial i\setminus j}\sum_{t_{k}^{'\left(k\right)}\geq t_{i}^{\left(j\right)}-1}m_{k\to\psi_{i}}\left(t_{i}^{\left(j\right)},t_{k}^{'\left(k\right)},g_{i}^{\left(j\right)}\right)\nonumber \\
 &  & - \; m_{i\to\psi_{i}}\left(t_{i}^{\left(j\right)},g_{i}^{\left(j\right)}\right)\1\left(t_{i}^{\left(j\right)}<t_{j}^{'\left(j\right)}+1\right)\prod_{k\in\partial i\setminus j}\sum_{t_{k}^{'\left(k\right)}>t_{i}^{\left(j\right)}-1}m_{k\to\psi_{i}}\left(t_{i}^{\left(j\right)},t_{k}^{'\left(k\right)},g_{i}^{\left(j\right)}\right)\nonumber
\end{eqnarray}
}where in \eqref{eq:pass} we use the fact that $\delta\left(t_{i},\left(1+\min_{j\in\partial i}\left\{ t_{j}^{'\left(j\right)}\right\} \right)\right)=\prod_{j\in\partial i}\1\left(t_{i}\leq t_{j}^{'\left(j\right)}+1\right)-\prod_{j\in\partial i}\1\left(t_{i}<t_{j}^{'\left(j\right)}+1\right)$.
The last equation \eqref{eq:pass} can be computed efficiently. Similarly,
\begin{eqnarray}
p_{\psi_{i}\to i}\left(t_{i},g_{i}\right) & \propto & \delta\left(t_{i},0\right)\prod_{k\in\partial i}\sum_{t_{k}^{'\left(k\right)}}m_{k\to\psi_{i}}\left(0,t_{k}^{'\left(k\right)},g_{i}\right)+\\
 &  & +\prod_{k\in\partial i}\sum_{t_{k}^{'\left(k\right)}\geq t_{i}-1}m_{k\to\psi_{i}}\left(t_{i},t_{k}^{'\left(k\right)},g_{i}\right)\nonumber\\
 &  & -\prod_{k\in\partial i}\sum_{t_{k}^{'\left(k\right)}>t_{i}-1}m_{k\to\psi_{i}}\left(t_{i},t_{k}^{'\left(k\right)},g_{i}\right)\nonumber
\end{eqnarray}

A more efficient parametrization of the equations can be derived by noting that in the right-hand side of \eqref{eq:pass}, incoming distributions $m_{i\to\psi_i}$ are aggregated in a simple way. In this way the BP update of $\psi_{i}$ can be computed in time $O\left(G\cdot T\cdot\left|\partial i\right|\right)$, where $G$ is the maximum allowed recovery delay, and the one of $\phi_{ij}$ in time $O\left(G^{2}\cdot T\right)$. In practice, $G$ can be taken constant for a geometric distribution $\mathcal G$.
A single BP iteration can be thus computed in time $O\left(T\cdot G^{2}\cdot\left|E\right|\right)$. We remark that the BP equations for the posterior distribution are exact (and have a unique solution) on tree factor graphs \cite{pearl_reverend_1982}. As the topology of the factor graph mirrors the one of the original graph, \eqref{eq:effective} allows the exact computation of posterior marginals for the SIR model on tree graphs (at difference with the DMP method).
\section{Bethe Free Energy}\label{sec:free-energy}
The cavity scheme allows one to compute approximately the Free Energy of the system $f = -\log Z$. In our case, $\log Z$ corresponds to the log-likelihood of external parameters, and thus can be used to estimate them. One of the many expressions of the Bethe free energy as a function of BP messages and their normalizations has the following form (see e.g. \citep{mezard_information_2009}):
\begin{equation}
	-f = \sum_a f_a + \sum_i f_i - \sum_{ia} f_{ia}\label{eq:free-energy}
\end{equation}
where
\begin{eqnarray}
f_a & = & \log \left( \frac{1}{|\partial a|}\sum_{i\in \partial a} z_{ai} \sum_{z_i} p_{F_a\to i}(z_i) m_{i \to F_a}(z_i) \right)\\
f_{ia} & = & \log \left(\sum_{z_i}p_{F_a\to i}(z_i) m_{i\to F_a}(z_i)\right)\\
f_i & = & \log \left(\sum_{z_i} \prod_{i\in a} p_{F_a\to i}(z_i)\right)
\end{eqnarray}

\section{Convergence of the BP equations}

In some cases the BP equations seem not to converge, or to require too large a number of iterations to converge. This is often (but not always!) the case when the information present in the observation is insufficient to perform good inference on the initial conditions. We found out that this limitation is not showstopping; in almost all cases very useful information is still present on the unconverged marginals. A simple strategy, consisting in averaging the probability of being the origin of the epidemics on a number of BP iterations (e.g. 100), gives excellent results in most cases. A second fact we observed in simulations is that the equations seem to converge almost always when the seed set is fixed; in those cases the estimation of the BP log-likelihood (the free energy of our model) helps to identify the correct origin much more precisely than with DMP. Of course, this add an overall factor $N$ to the algorithm because the simulation must be performed for each possible seed.

\section{Dynamic Message Passing}
Dynamic message passing (DMP) \citep{lokhov_inferring_2013} attempts to infer the zero patient in the following way. First, Bayes' theorem suggests that the desired probability $P(i|\mathbf{x^T})$ of the seed being site $i$ given an observation $\mathbf x^T$ of an epidemic at time $T$, is proportional to $P(\mathbf{x^T}|i)$. Then DMP considers an approximation of this latter probability in a factorized form:
\begin{eqnarray}
P(\mathbf{x^T} | i) &\simeq& \prod_k P^k({x_k^T}|T,i)\\
\label{eq:DMP} 	 & =& 	\prod_{k| x_k^{T}=S} P_S^k(T,i) \prod_{l|x_l^{T}=I} P_I^l(T,i) \prod_{m|x_m^{T}=R} P_R^m(T,i). \label{factorization}
\end{eqnarray}
In this expression $P_S^k(T,i)$  ($P_I^l(T,i)$, $P_R^m(T,i)$) is the probability that site $k$ ($l,m$) is found in the susceptible (infected, recovered) state at observation time $T$ if the epidemic started at site $i$. Note that $P_S^k(T,i)$, $P_I^l(T,i)$ and $P_R^m(T,i)$ are probabilities in the subindex $S,I,R$ and not in $i$ or $T$. In the first line we keep the notation used in \cite{lokhov_inferring_2013}.

The actual values of  $P_S^k(T,i)$, $P_I^l(T,i)$ and $P_R^m(T,i)$ are obtained by iterating the forward propagation equations \citep{karrer_message_2010} of the SIR model, starting from node $i$ in the graph. For each possible origin an energy-like function can be defined as $E(i) = -\log P(\mathbf{x^T} | i)$, and the most probable seed is therefore the one with lowest energy.

In the standard dynamic message passing approach \cite{lokhov_inferring_2013}, the forward equations for each possible origin are propagated over all nodes in the graph, to compute the terms $P_S^k(T,i)$, $P_I^l(T,i)$ and $P_R^m(T,i)$. This includes those nodes that are not participating in the actual epidemic (susceptible nodes at observation time).
A restricted version of DMP can be implemented by only iterating the equations \citep{karrer_message_2010} over the connected component of nodes that do participate in the observed epidemic. This means that any node of the graph that is susceptible and is surrounded by susceptible nodes at observation time $T$ is effectively removed from the DMP equations. In this way the contributions to the probabilities $P_S^k(T,i)$, $P_I^l(T,i)$ and $P_R^m(T,i)$ come only from epidemic paths that are consistent with the observed epidemic. In other words, the effective graph for the epidemic transmission is only the part of the original graph that got infected, and its boundary. The susceptible boundary is not used in the message passing procedure, but it is evaluated in the first factor of equation \ref{eq:DMP}. In the case in which there are unobserved nodes, they are logically considered as possible infected/recovered nodes and, therefore, the DMP equations are iterated over the connected
component of infected, recovered and unobserved nodes. We found that this approach, that we will call Restricted Dynamic Message Passing (DMPr), in many situations gives better estimates, as is shown in Fig. 4 of the manuscript, and also in figures \ref{fig:T6mu1}, \ref{fig:T11mu1} and \ref{fig:T11mu05}.

Despite this improvement, DMPr is still less accurate than BP. Figs.~\ref{fig:T6mu1}, \ref{fig:T11mu1} and \ref{fig:T11mu05} show three sets of parameters as benchmarks. In all cases the observation time is known, and epidemics were generated with only one seed. We found that the belief propagation equations sometimes do not converge when the epidemic becomes large compared to the size of the graph (see Figs. \ref{fig:T11mu1} and \ref{fig:T11mu05}). For small observation times ($T=5$ in Fig. \ref{fig:T6mu1}) the epidemic remains small and BP always converges. The lack of convergence is associated to a decrease in accuracy and it is in this region where our algorithm performs (slightly) worse. It is important to underline that the amount of information concerning the origin of an epidemic is reduced when the epidemic covers almost all the supporting graph, and not surprisingly all algorithms perform worse in this region (as in Fig.  \ref{fig:T11mu1}).

\begin{figure}
  \includegraphics[width=8cm]{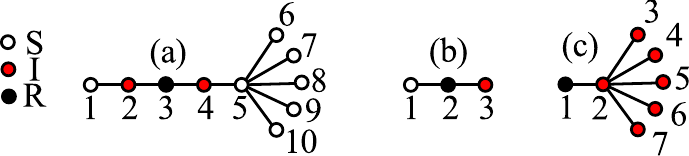}
  \caption{Examples of propagations in which the factorization assumption in (\ref{factorization}) leads to incorrect results. For (a) the transmission probability on each edge is $\lambda=0.5$, the recovery probability for each vertex is $\mu=0.3$ and the observation time is $T=8$; for (b) $\lambda=0.4$, $\mu=0.1$ and $T=5$; for (c) $\lambda=0.2$, $\mu=0.2$ and $T=8$. }
  \label{Fig_Spazzolone}
\end{figure}

\begin{figure}
\includegraphics[width=0.3\columnwidth, angle=270]{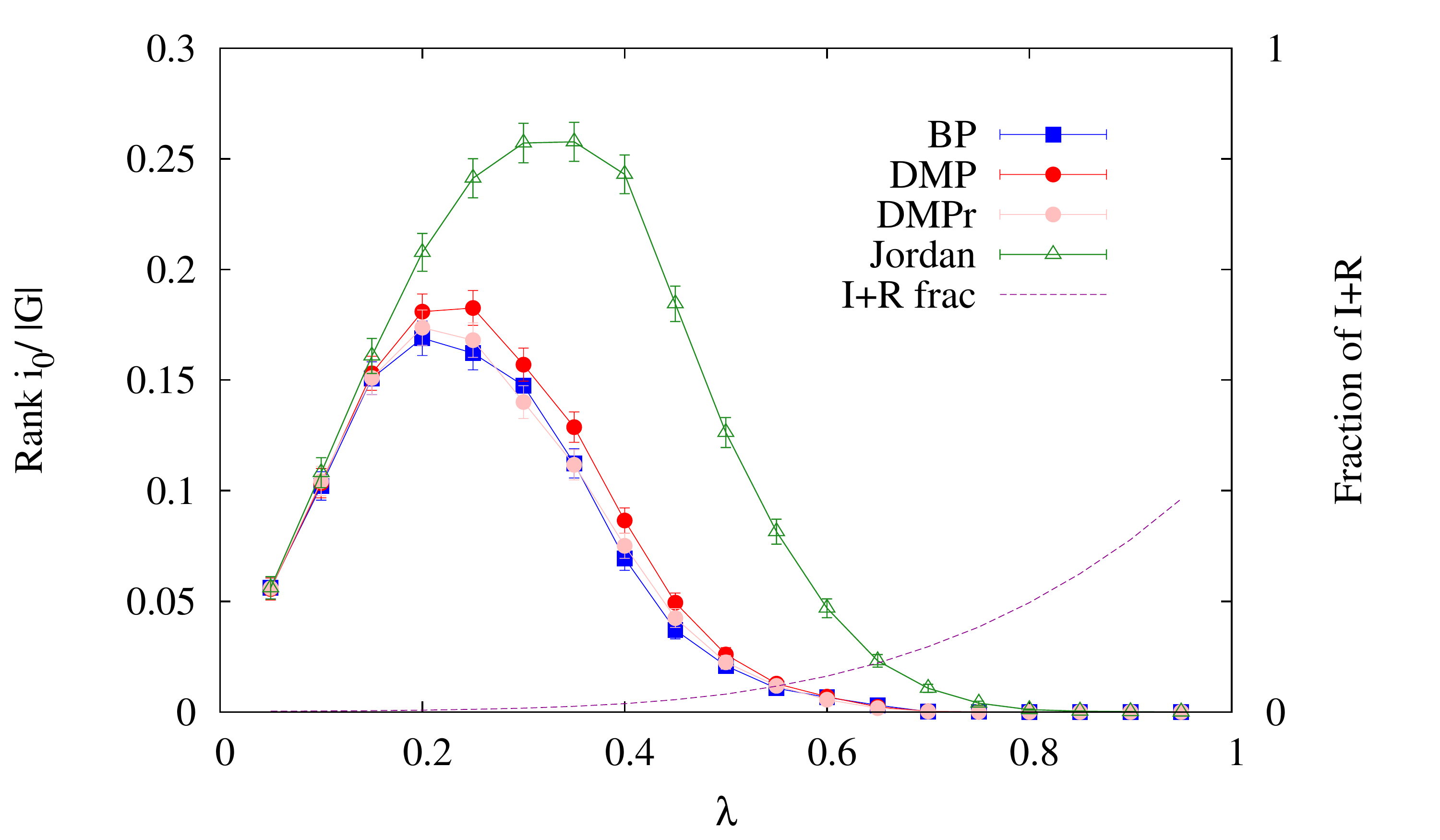}
\includegraphics[width=0.3\columnwidth, angle=270]{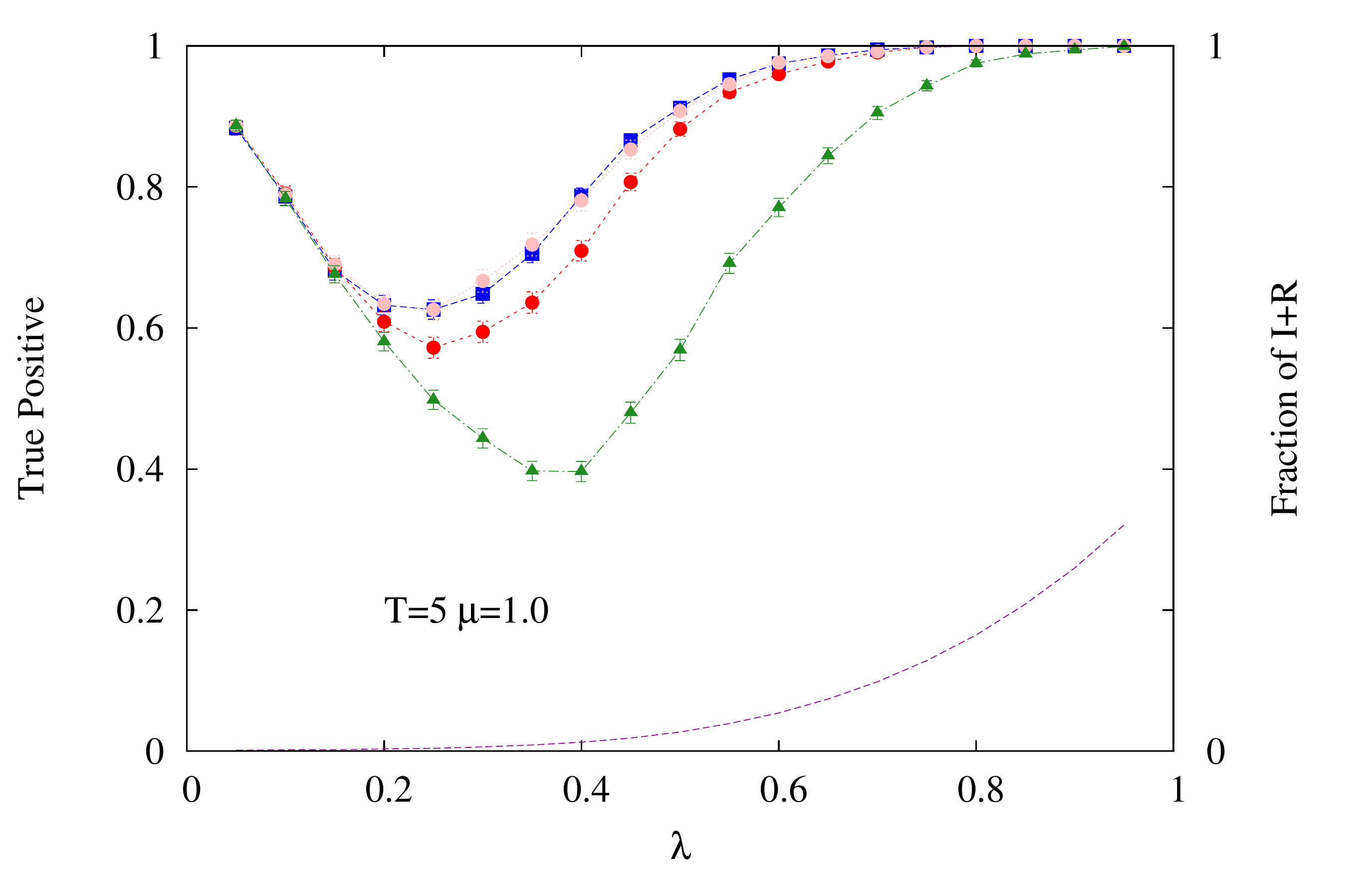}
\caption{{\bf Left:} Relative ranking of the true zero patient with respect to the size of the epidemic, as a function of the transmission probability $\lambda$. The size of the epidemic is shown as a dashed line, and corresponds to the right y axis. {\bf Right:} Fraction of the instances in which the true zero patient is found by each algorithm. The forward epidemic is propagated until observation time $T=5$, with recovery probability $\mu=1$. Simulations were run over 1000 samples of random regular graphs with $N=1000$ and degree 4.}
\label{fig:T6mu1}
\end{figure}

\begin{figure}
\includegraphics[width=0.3\columnwidth, angle=270]{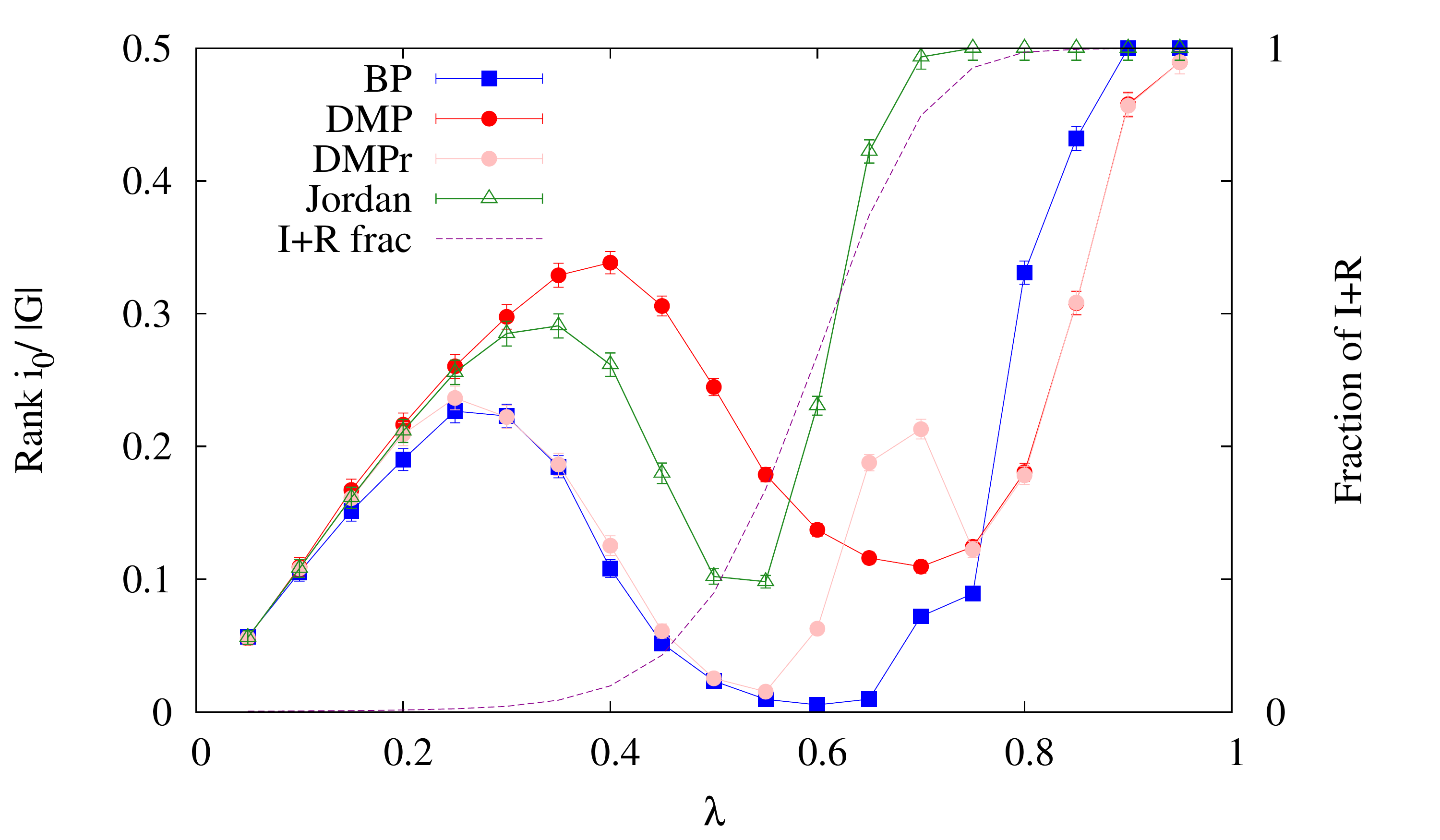}
\includegraphics[width=0.3\columnwidth, angle=270]{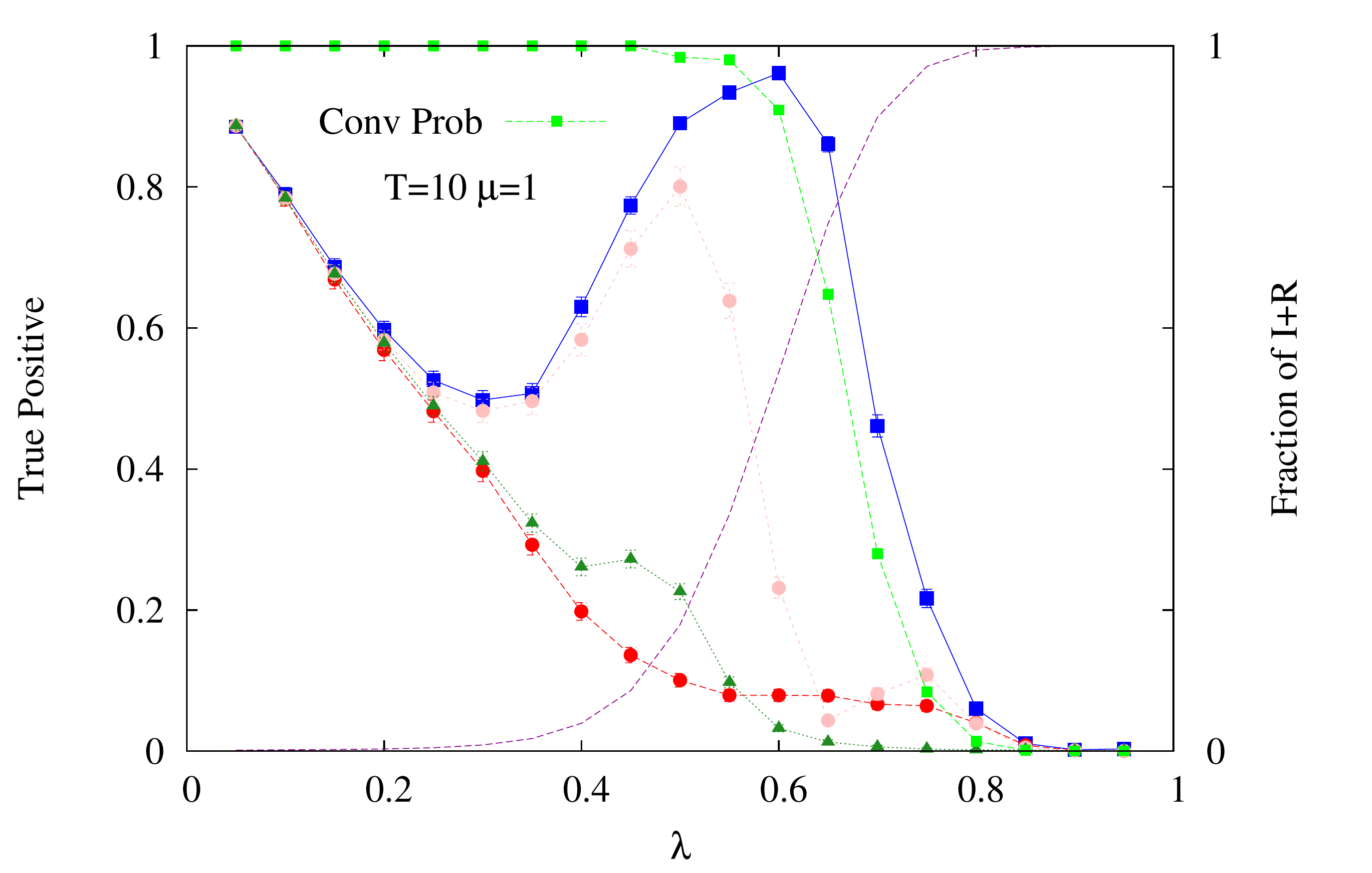}
\caption{{\bf Left:} Relative ranking of the true zero patient with respect to the size of the epidemic, as a function of the transmission probability $\lambda$. The size of the epidemic is shown as a dashed line, and corresponds to the right y axis. {\bf Right:} Fraction of the instances in which the true zero patient is found by each algorithm. The forward epidemic is propagated until observation time $T=10$, with recovery probability $\mu=1$. Simulations were run over 1000 samples of random regular graphs with $N=1000$ and degree 4.}
\label{fig:T11mu1}
\end{figure}

\begin{figure}
\includegraphics[width=0.3\columnwidth, angle=270]{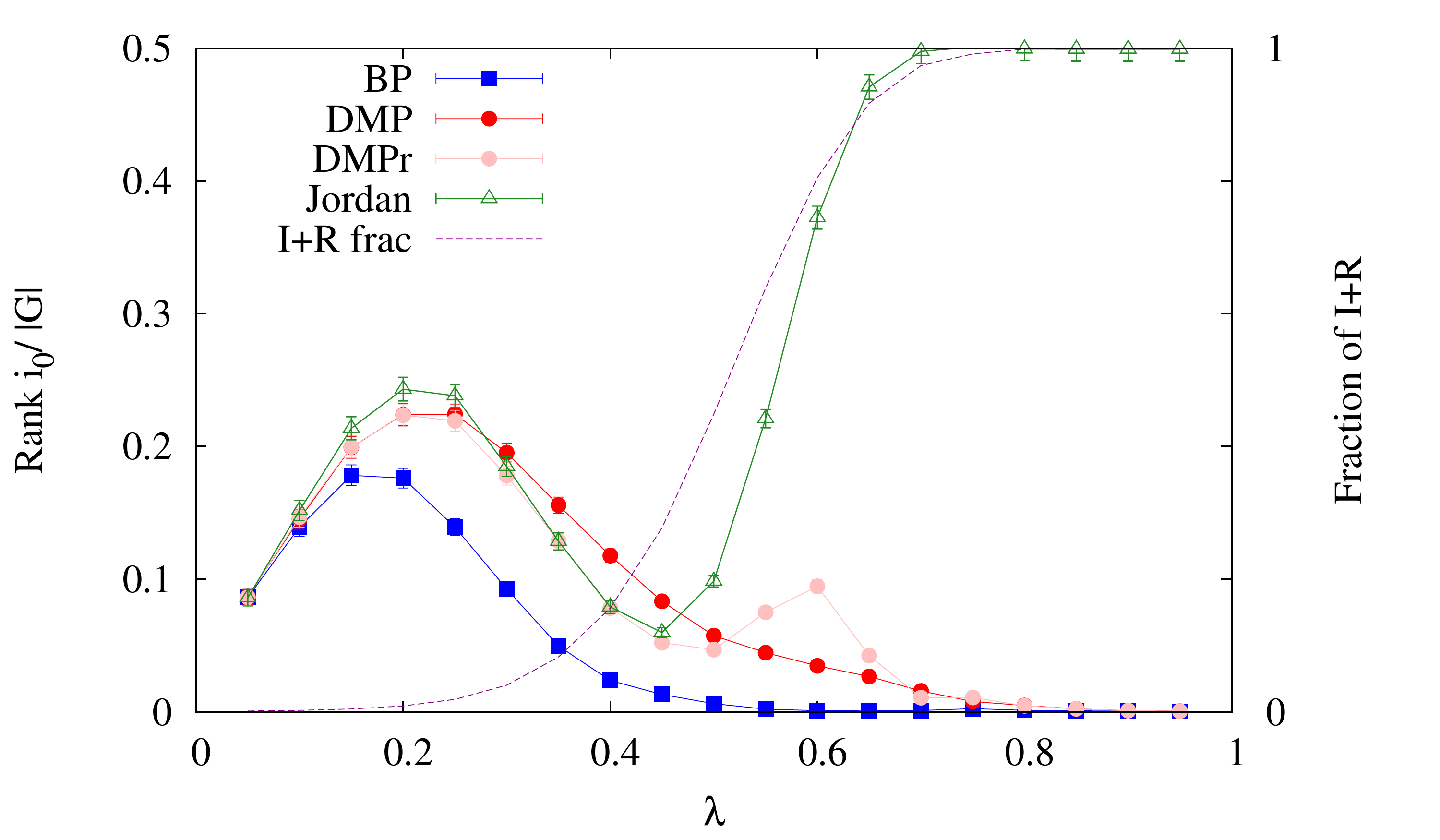}
\includegraphics[width=0.3\columnwidth, angle=270]{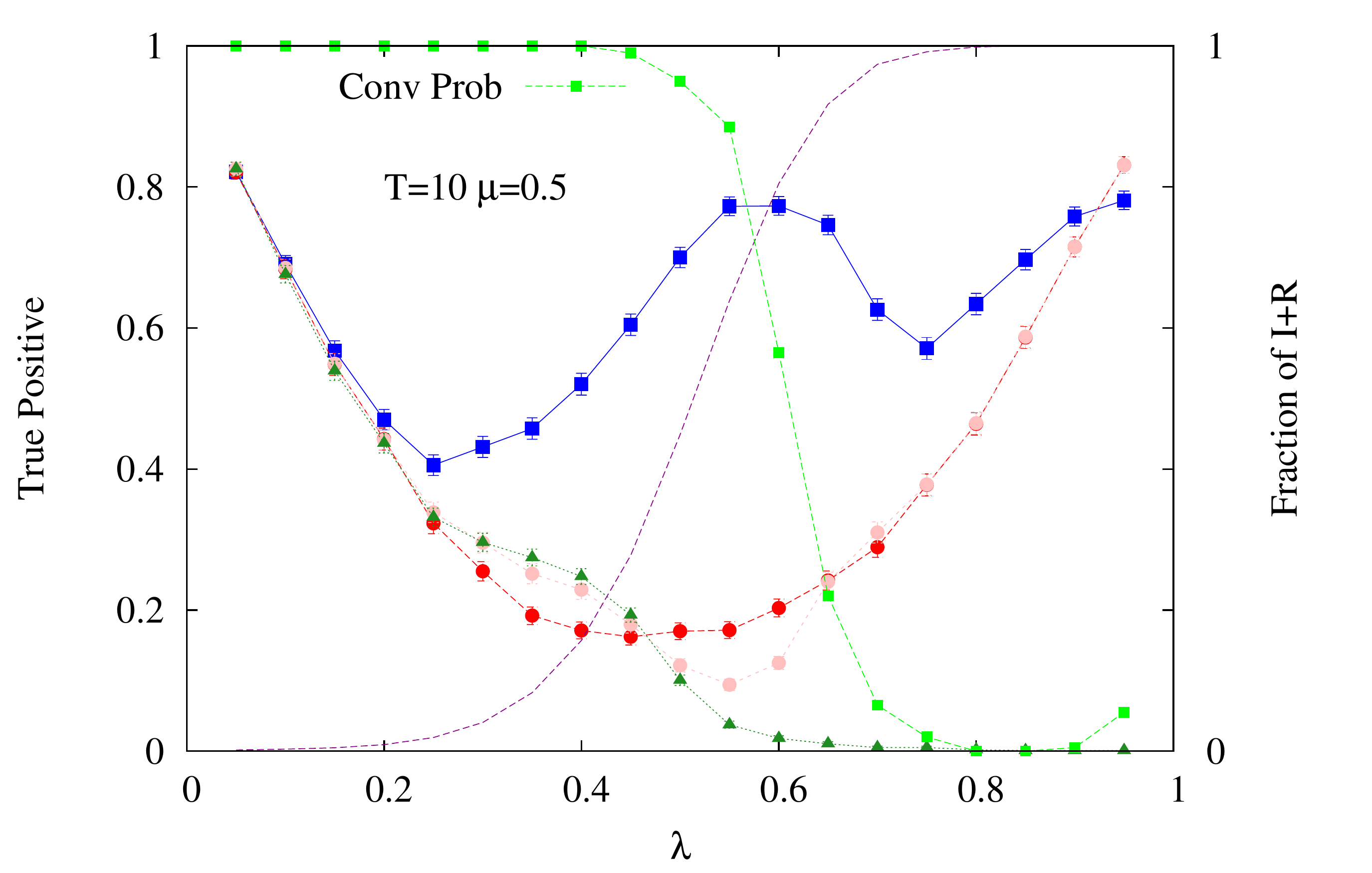}
\caption{{\bf Left:} Relative ranking of the true zero patient with respect to the size of the epidemic, as a function of the transmission probability $\lambda$. The size of the epidemic is shown as a dashed line, and corresponds to the right y axis. {\bf Right:} Fraction of the instances in which the true zero patient is found by each algorithm. The forward epidemic is propagated until observation time $T=10$, with recovery probability $\mu=0.5$. The figure in the right corresponds to the one shown in the manuscript as Fig 2, but with a different parameterization of the x axis.  Simulations were run over 1000 samples of random regular graphs with $N=1000$ and degree 4.}
\label{fig:T11mu05}
\end{figure}

It is important to notice that the factorization in (\ref{factorization}) is an approximation which can lead to significant errors even when the contact network is a tree (and therefore the solution of the time-forward propagation equations is exact). As an extreme example one can consider the case shown in Fig.~\ref{Fig_Spazzolone}a. In this case, the most probable position of the seed is at 3, while DMP places the seed at 2. An intuitive explanation of this error is the following: the DMP energy contains a number of contributions from the leaves 6--10 which are larger when the seed is placed at 3 than when it is placed at 2 (because the probability that any of the vertices 6--10 remains \emph{S} increases with their distance from the seed). This contribution should in fact be ignored, because the probability that the leaves 6--10 are \emph{S} conditioned on the node 5 being \emph{S} is 1 independently on the position of the seed, and we know that vertex $5$ does not get infected. The factorization assumption in (\ref{factorization}) ignores this. If the number of leaves is large enough, their contribution dominates the energy, and the position at which the seed is placed by DMP is ``pushed'' to the left. Moreover, if the chain 1--5 is longer, the position at which the seed is placed by DMP can be up to the left-most \emph{I} seed, and the rank of the actual seed (at the center of the chain) can be arbitrarily large. This example is correctly solved by DMPr,  which ignores the \emph{S} leaves to the right of vertex 5. A smaller example based on the same principle is given in Fig. \ref{Fig_Spazzolone}b; in this case also DMPr gives a wrong answer, as it is unable to eliminate any node. An example with no susceptible nodes is given in Fig. \ref{Fig_Spazzolone}c. The most probable seed position is node 1, whereas DMP and DMPr predict it to be node 2. Here each of the leaves 6--10 in DMP/DMPr ``attracts'' the seed towards themself, even though for $T$ large enough and conditioned to the fact that node 2 is infected its real contribution to decide between nodes 1 and 2 is negligible, and the most likely position of the seed is mostly influenced by the fact that node 1 is the only \emph{R} node.

\section{Evolving networks}
We studied the case of dynamically evolving networks, in which the transmission probility $\lambda_{ij}$ depends on time (representing the time-evolution of interactions between agents).
To cope with time dependent transmission probabilities, it suffices to consider transmission delay probabilities that depend on $t_i$, i.e.
$\omega_{ij}(s_{ij}|g_i,t_i)=\lambda_{ij}(t_i+s_{ij})\prod_{t=0}^{t_i-1} (1-\lambda_{ij}(t+s_{ij}))$ for $s_{ij} \leq g_i$ and $\omega_{ij}(\infty|g_i,t_i)=1-\sum_{s=0}^{g_i}\omega_{ij}(s|g_i,t_i)$, the rest of the formalism remaining the same.

We considered two interesting datasets, each one consisting in a large list of time-stamped contacts between pairs of individuals, which we aggregated into $\Delta T$ effective time steps. We focused on the known initial time scenario $T_0=0$ to facilitate the comparison between algorithms. We simulated the progression of many virtual epidemics initiated by single random individuals.
\subsection{Proximity contacts network}
\begin{figure}
\includegraphics[width=0.45\textwidth]{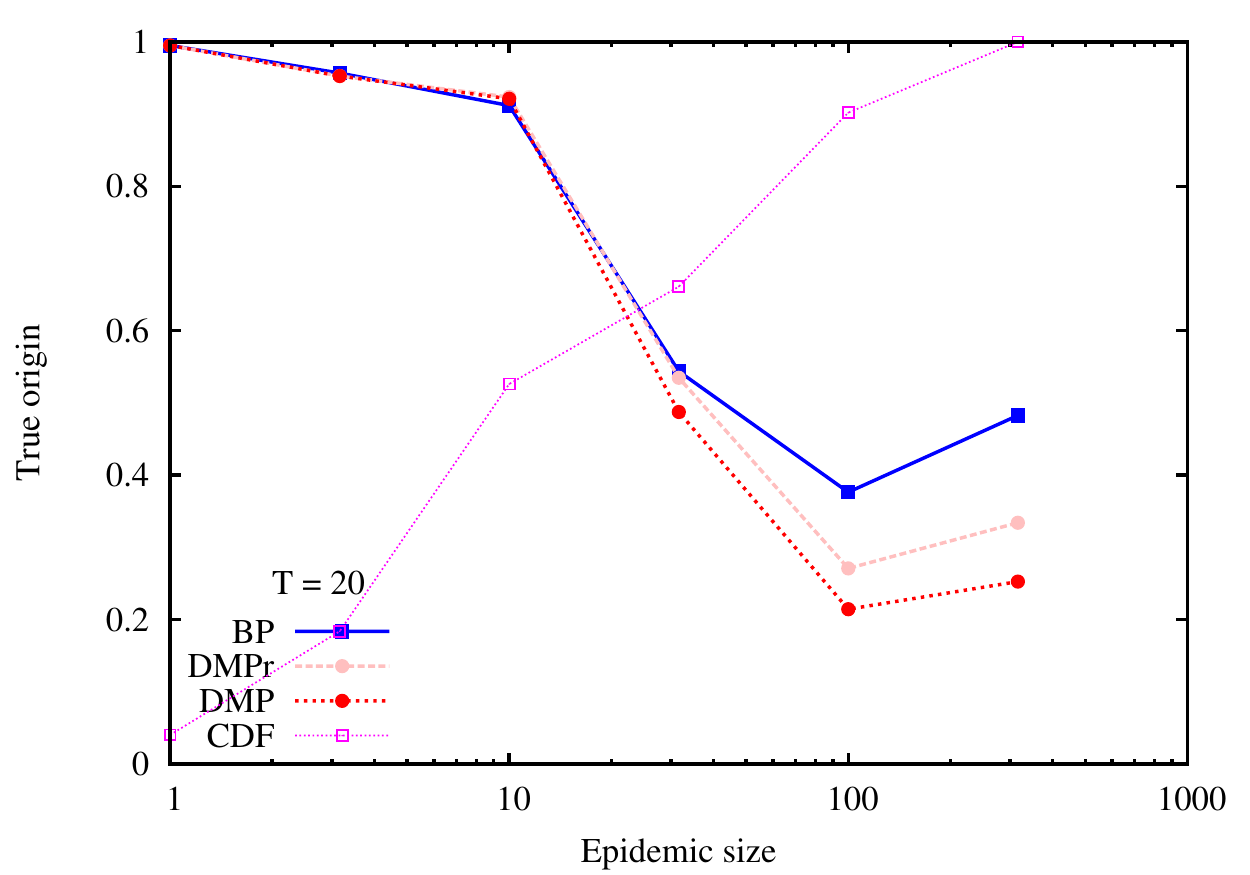}\includegraphics[width=0.45\textwidth]{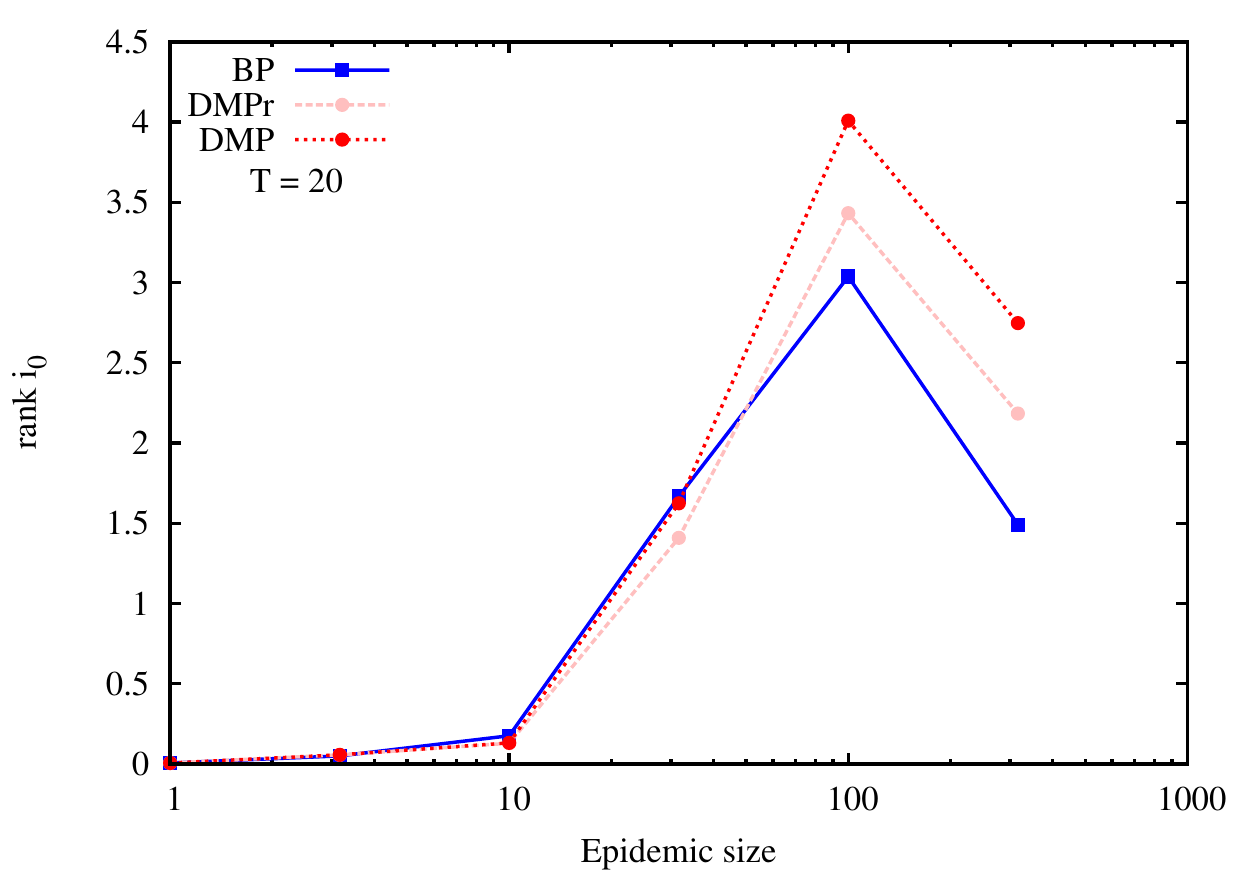}

\includegraphics[width=0.45\textwidth]{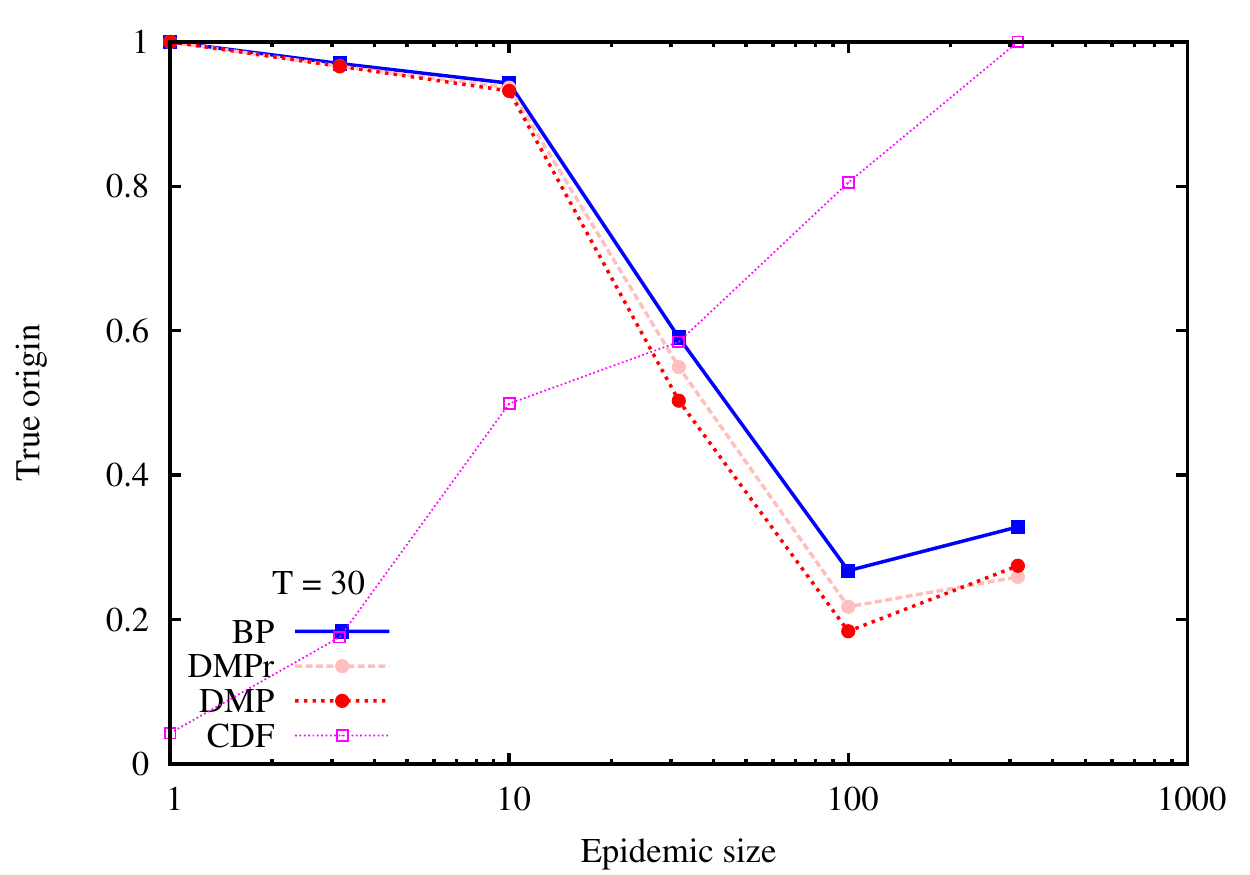}\includegraphics[width=0.45\textwidth]{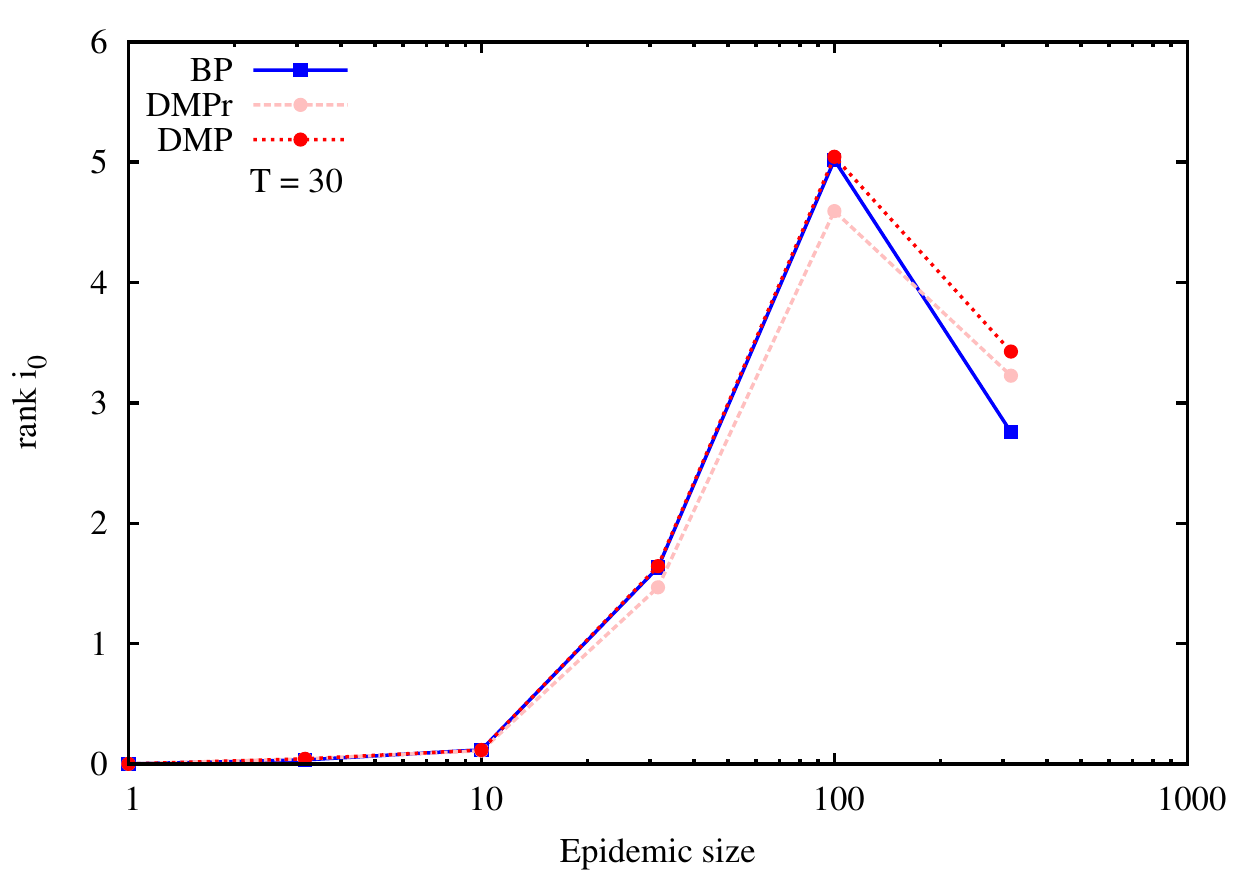}
\caption{Performance of BP, DMPr and DMP on the proximity contacts temporal network.\label{fig:contacts}. The cumulative distribution function (CDF) of the size for the 5000 epidemics for $T=20$ (resp. 2000 epidemics for $T=30$) is plotted on the left panels.}

\end{figure}

The first dataset \cite{isella_whats_2011} corresponds to $20$ seconds face-to-face contacts in an exhibition, obtained using badges with RFID technology. Here the case $T_0=0$ is particularly significant because it corresponds to the situation in which the zero patient was infected before entering the facilities. We concentrated on the day in which the number of individuals and contacts was larger. We employed the following set of parameters: probability of contagion in a 20 second interval $\lambda^{20s}_{ij}=0.2$, recovery probability $\mu^{20s}=0.0014$. We selected the day in the dataset in which the number of individuals and interactions was largest. In Fig.~\ref{fig:contacts} we illustrate the results on a large number of random virtual epidemics. We simulated 5000 virtual epidemics for $\Delta T=20$ (resp. 2000 virtual epidemics for $\Delta T=30$) initiated by a random individual. In each case, nodes for which being the seed was impossible due to topological constraints were identified in a fast preprocessing. We then attempted the inference of the zero-patient with BP, DMP and DMPr.

\subsection{Sexual contacts network}
\begin{figure}
\includegraphics[width=0.45\textwidth]{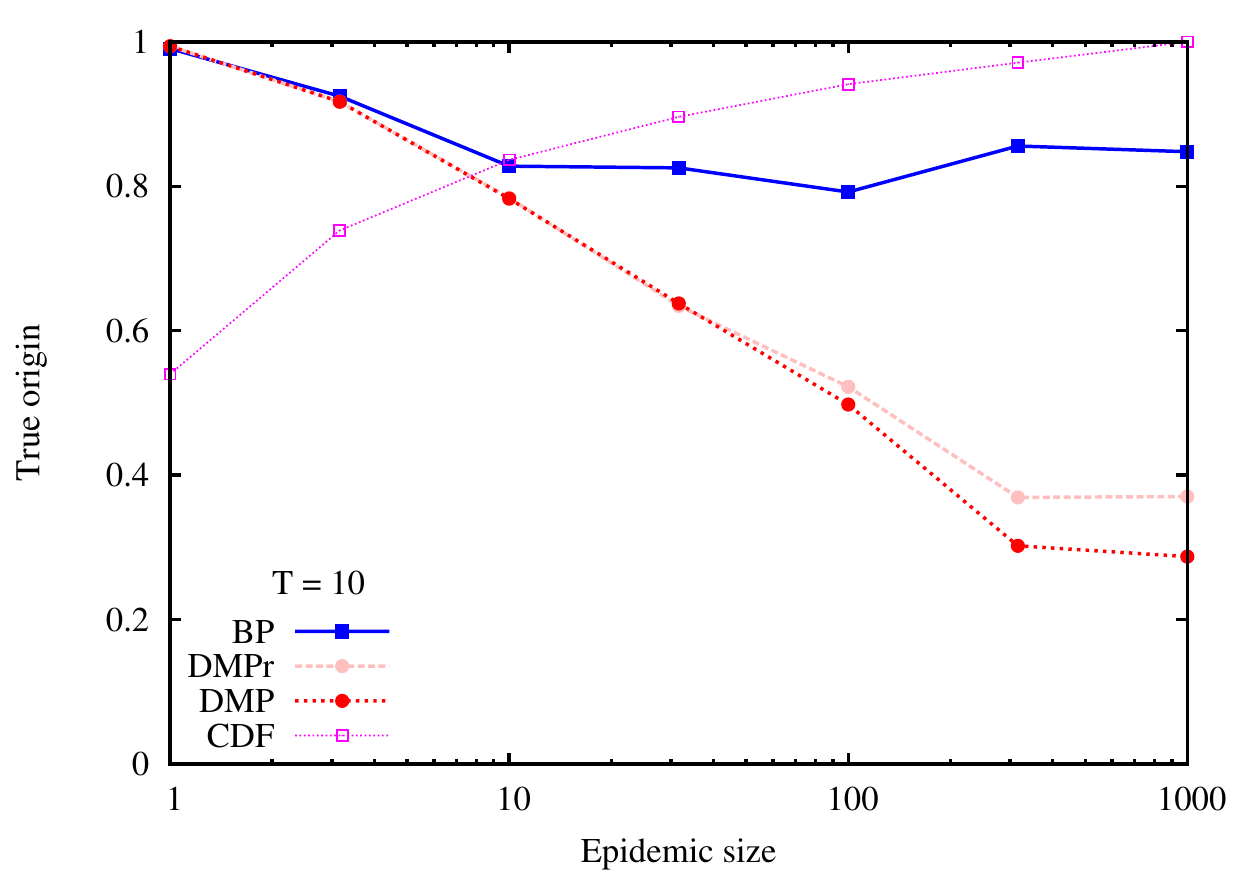}\includegraphics[width=0.45\textwidth]{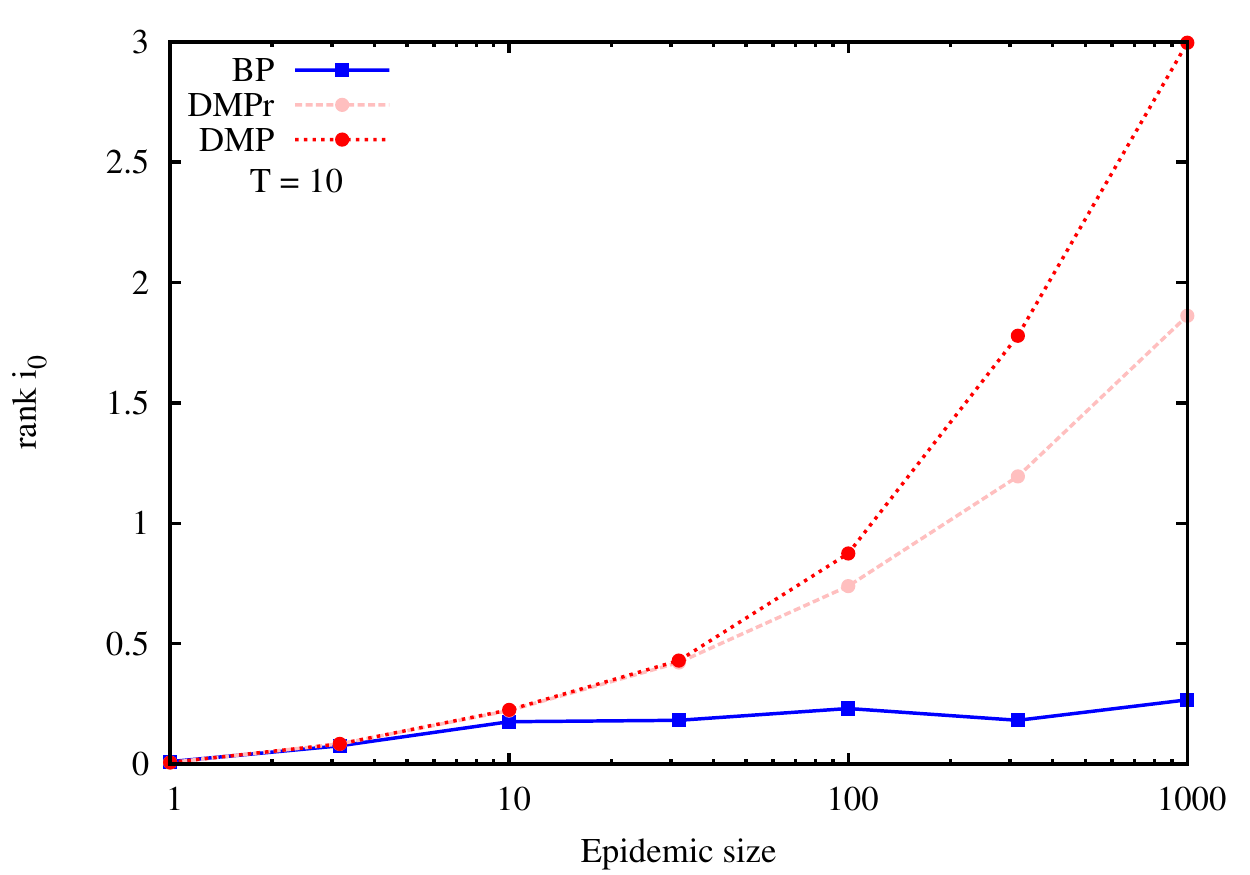}

\includegraphics[width=0.45\textwidth]{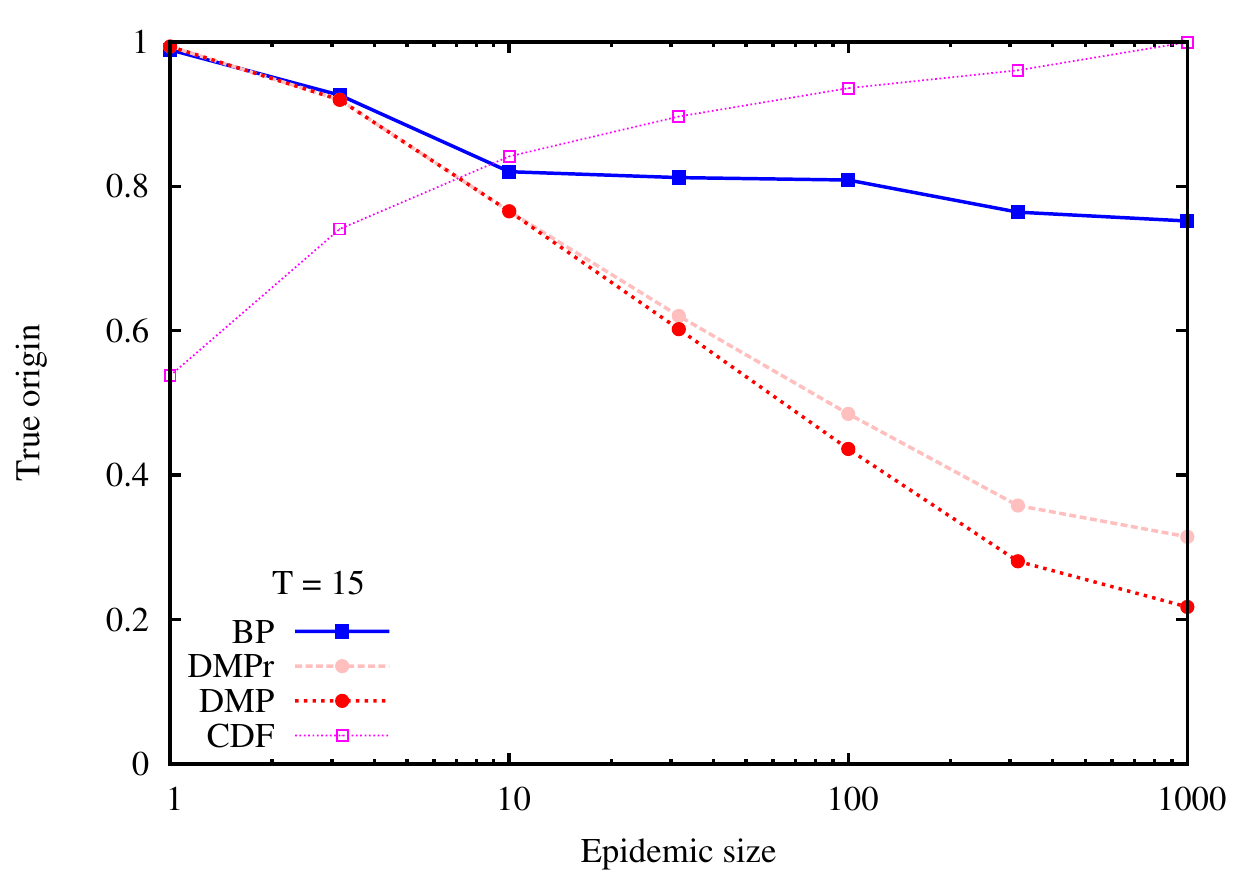}\includegraphics[width=0.45\textwidth]{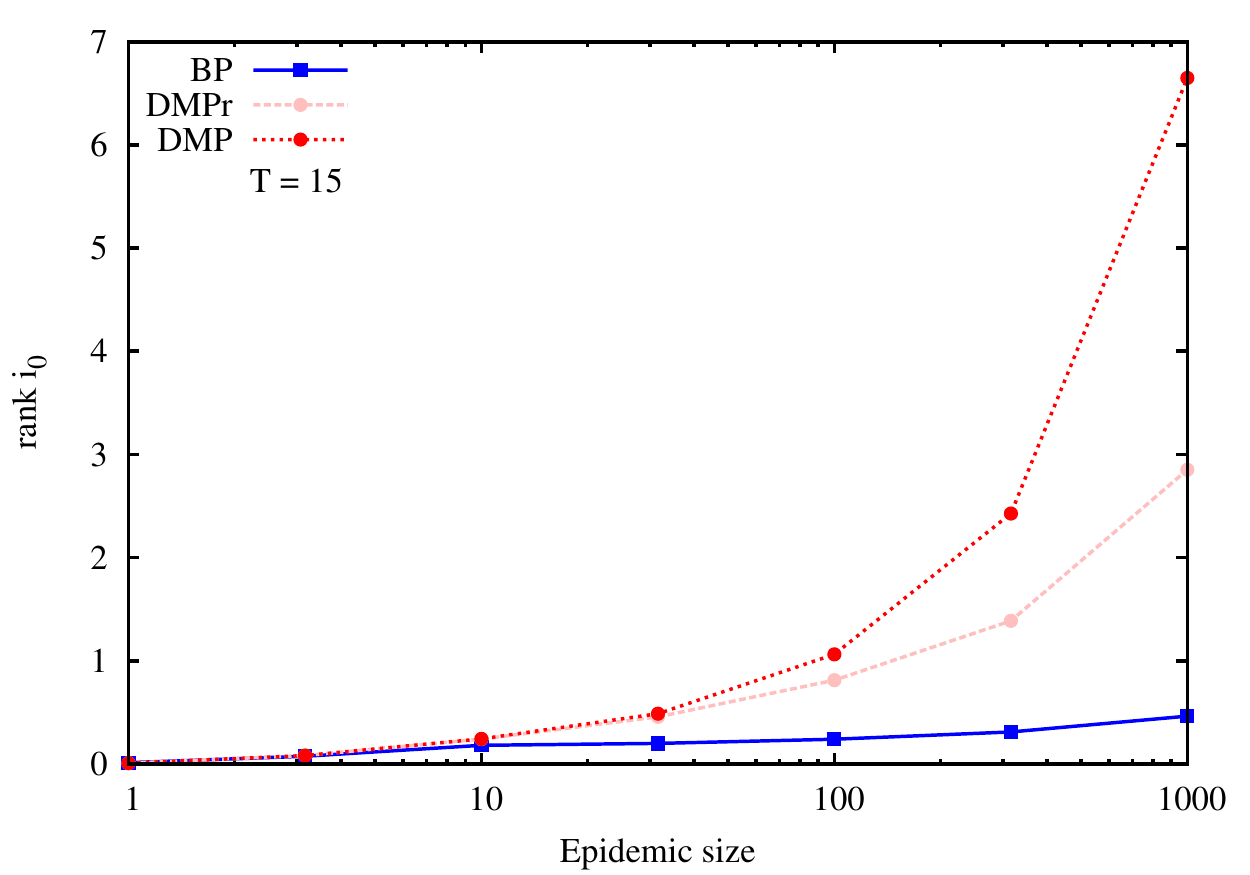}

\includegraphics[width=0.45\textwidth]{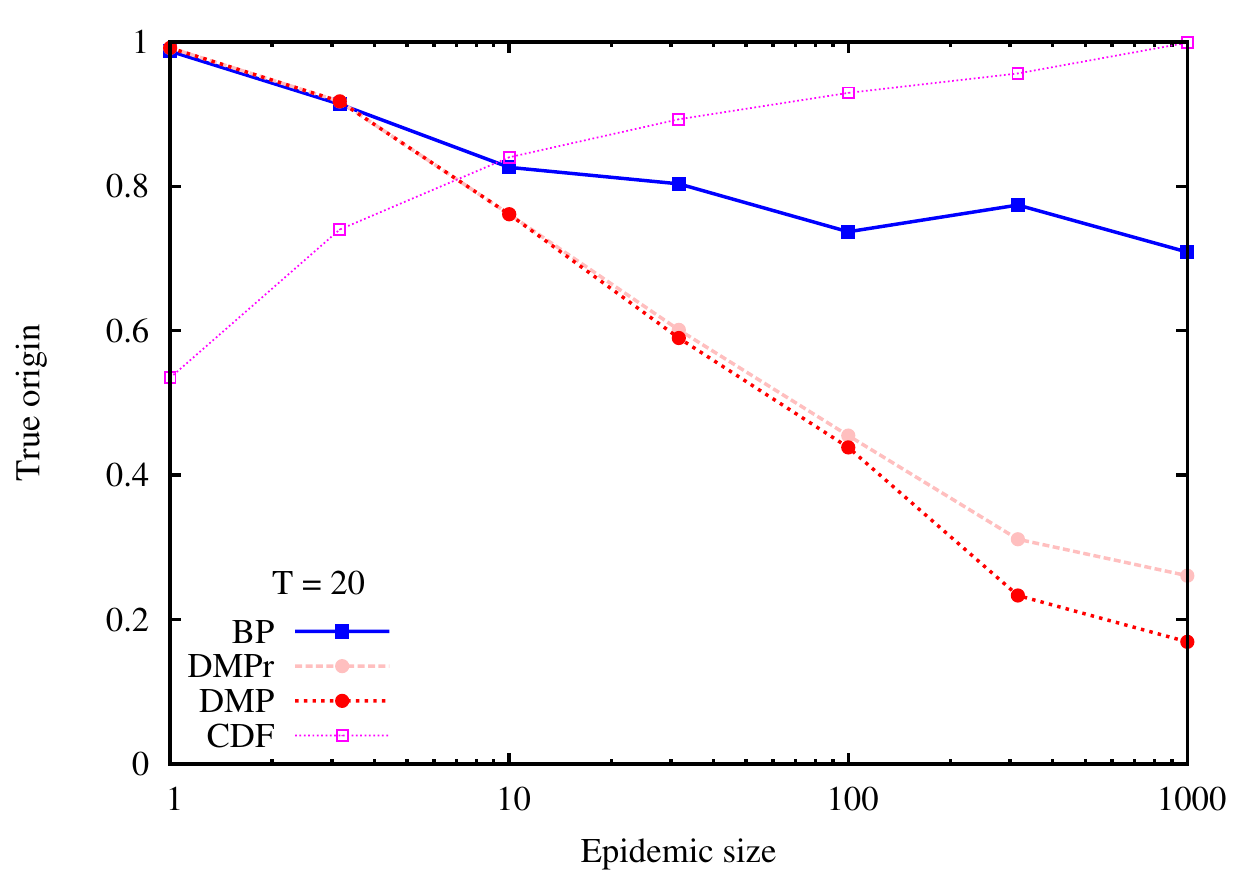}\includegraphics[width=0.45\textwidth]{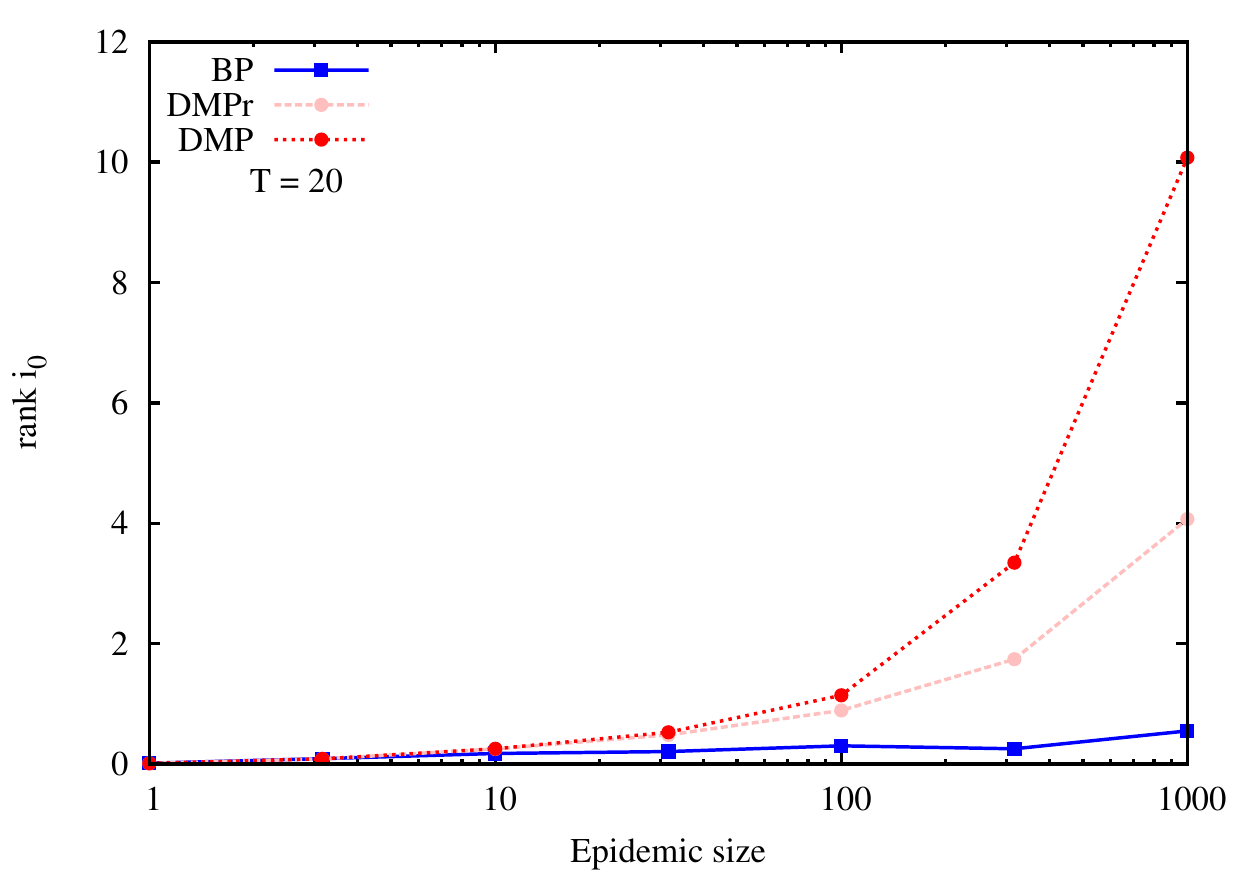}
\caption{Performance of BP, DMPr and DMP on the sexual contacts temporal network.\label{fig:escorts} The cumulative distribution function (CDF) of the size for the 10\,000 epidemics for each $T$ is plotted on the left panels}
\end{figure}

The second dataset \cite{rocha_information_2010} is a database of sexual encounters between clients and escorts self-reported by clients on a Brazilian website. We fixed the probability of transmission in a single contact as $\lambda^{contact}_{ij}=0.2$ and the yearly recovery probability as $\mu^{year}=0.5$. We selected the records of the last two years (slightly over half of the dataset), because the oldest data is sparser and seems relatively incomplete (and leads to very small epidemics). The results are summarized in Fig.~\ref{fig:escorts} and Table I of the manuscript.

We also tried to assess the relevance of temporal correlations in the efficiency of the inference processes. In particular, we replicated the analysis above in a different regime of parameters in which temporal correlation was shown to have a significant effect in the size of the outbreak \cite{rocha_simulated_2011}. After reshuffling the edges of the network and reassigning the contact times randomly (as in the method ``Configuration Model'' of \cite{karsai_small_2011} but preserving the bipartite structure of the network). Specifically, we employed a time-window of the last 800 days with deterministic recovery times at 90 days and infectivity parameter $\lambda=0.7$. This is a regime of parameters in which it was shown \cite{rocha_simulated_2011} that the ''Configurational Model'' reshuffled version of the network leads to significantly smaller outbreak size than the original one. We used $\Delta T=80$ and we obtained indeed that outbreaks involved $437\pm82$ individuals for the original dataset and $258\pm58$ for the reshuffled one over 500 samples. However, the impact on the inference performance was limited in this case, the probability of inferring the correct zero patient was $93\%\pm1\%$ against $94\%\pm1\%$ over 500 samples, $73\%\pm4\%$ against $75\%\pm4\%$ when restricted to samples with more than 10 infections. More work would be needed to clarify this aspect with more generality.

\section{Inference with multiple seeds}

\begin{figure}
\includegraphics[width=0.5\columnwidth, angle=0]{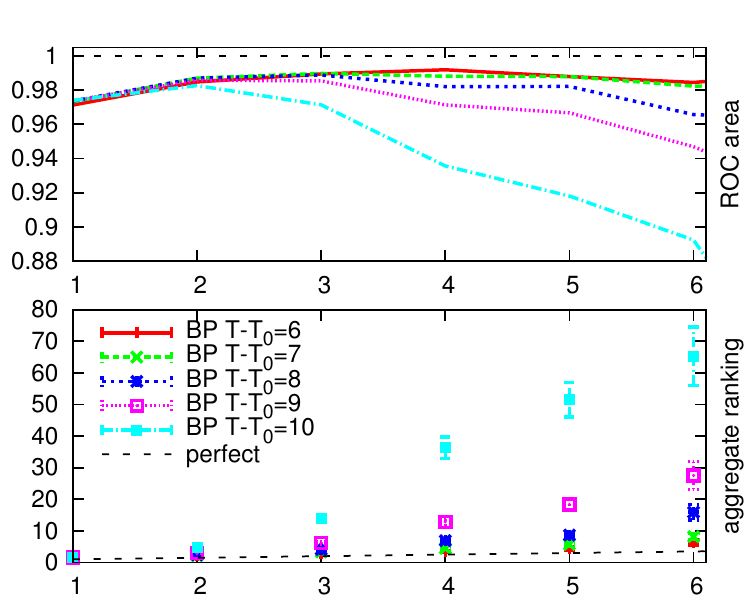}
\caption{Inference of multiple seeds on 1 000 virtual epidemics with $\mu=1,\lambda=0.5$, seed probability $0.002$ and different observation times $T-T_0$ (unknown to the inference algorithm). Bottom: average sum of the rankings of the true seeds in the inferred order is plotted vs. the true number of seeds. Top: the average normalized mean ranking, that equals the average area of the ROC curve. Each ROC curve was computed restricted to the epidemic subgraph.\label{fig:multiseeds-sm} }
\end{figure}

If the epidemics initiates at multiple seeds, methods based on the exhaustive exploration of initial states like DMP suffer a combinatorial explosion; the straight forward generalization of DMP would require $\binom{N}{k}$ different forward propagations to explore all subsets of $k$ seeds in a graph with $N$ vertices. This problem does not affect BP, as the trace over initial conditions is performed directly within the framework. A posterior distribution of seeds is obtained with a finite (non-infinitesimal) prior. We performed simulations by generating epidemic spreads with a (known) uniform prior seed probability. Note that this method does not provide the subset of maximum posterior probability but gives the probability of a single node to belong to a seed subset. As before, nodes are sorted by their seed probability and a ROC curve can be computed. Note that the area above the ROC curve is a natural generalization of the normalized rank for a single origin. An alternative generalization is the average of the ranks of the true origins, with the disatvantage that is bounded below by $\frac12 (k-1)$. A comparative plot of both measures is given in Fig.~\ref{fig:multiseeds-sm}

\section{Inference of epidemic parameters with BP and DMP}

Among the methods tested in this Letter, Jordan centrality is the least effective, but is also the least demanding in terms of prior information. As presented so far, both Dynamic Message Passing and Belief Propagation require not only the network of contacts, but also the values of  the transmission probability $\lambda_{ij}$, and the recovery probability $\mu_i$. At least in the case of homogeneous parameters $\lambda_{ij}\equiv\lambda,\mu_i\equiv\mu$, this requirement can be relaxed since both methods can be used to infer the epidemic parameters by comparing the energy or the free energy obtained for different pairs $(\lambda,\mu)$.
The probability of parameters $(\lambda,\mu)$ given an observation of an epidemic is
\begin{equation}
P\left( \lambda,\mu|\mathbf{x^T} \right) = \frac{P(\mathbf{x^T} |  \lambda,\mu) P( \lambda,\mu)}{P(\mathbf{x^T})} \propto P(\mathbf{x^T} |  \lambda,\mu)
\end{equation}
assuming a uniform prior for $(\lambda,\mu)$.

\subsection*{Dynamic Message Passing}
The DMP method gives a probability to the observed epidemic starting from every single possible seed, $P(\mathbf{x^T} | i, \lambda,\mu)$  (see \eqref{eq:DMP}), so we can write
\begin{equation}
P(\mathbf{x^T} |  \lambda,\mu) = \sum_i P(\mathbf{x^T} | i, \lambda,\mu)
\end{equation}
An estimate of the negated log-likelihood of the parameters can therefore be given as the DMP free energy
\begin{equation}
f_{\text{DMP}}(\lambda,\mu) = - \log P\left( \lambda,\mu|\mathbf{x^T} \right) \simeq -\log \sum_i P(\mathbf{x^T} | i, \lambda,\mu).
\end{equation}
We call this a free energy since the last expression can be written as  $ -\log \sum_i \exp(-E(i))$. The lower this free energy, the more likely the parameters.

\subsection*{Belief Propagation}
Following the approach of the present Letter, the posterior of the parameters can be computed as

\begin{equation}
\mathcal{P}\left(\lambda,\mu|\mathbf{x}^{T}\right)\propto\sum_{\mathbf{t,g,x^0}} \mathcal{P}\left(\mathbf{x}^{T}|\mathbf{t},\mathbf{g}\right)\mathcal{P}\left(\mathbf{t},\mathbf{g}|\mathbf{x}^{0}\right)\mathcal{P}\left(\mathbf{x}^{0}\right).
\end{equation}
where the trace goes over the infection times $\mathbf t$, the recovery delays $\mathbf g$ and the different choices of zero patients $\mathbf{x}_0$.

Therefore the desired log-likelihood of the parameters can be computed as the negated free energy of the model in \eqref{eq:effective}:
\begin{equation}
f(\lambda,\mu) = -\log Z
\end{equation}
where $Z$ is given in  eq. (\ref{eq:effective}). The exact calculation of $Z$ is computationally hard, but the same Bethe approximation used for the inference of $\mathbf{x}^0$ yields an approximate $f_{\text{BP}}(\lambda,\mu)$, given in \eqref{eq:free-energy} for the free energy of this model.

In the manuscript is presented an example of the functions $f_{\text{DMP}}(\lambda,\mu)$ and $f_{\text{BP}}(\lambda,\mu)$ for an epidemic observed at time $T=8$. The 2-dimensional space of parameters $(\lambda,\mu)$ is discretized and the (standardized) values of the free energies shown in a heat-plot. Note that every single point of the graph corresponds to a run of DMP and DMPr with the corresponding parameters.

\bibliographystyle{apsrev4-1}
\bibliography{inference}
\end{document}